\documentclass{iopjournal}

\usepackage{siunitx}
\usepackage{subcaption}
\usepackage{graphicx}
\usepackage{hyperref}
\begin{document}

\articletype{Paper} 

\title{A pushing–pulling captive bubble method for repeatable measurement of dynamic contact angles underwater}

\author{Koki Iwasaki$^{1,*}$\orcid{0009-0001-9737-752X}, Hiroyuki Ebata$^1$\orcid{0000-0002-5370-631X} and Hiroaki Katsuragi$^1,$\orcid{0000-0002-4949-9389}}

\affil{$^1$Department of Earth and Space Science, The University of Osaka, Osaka, Japan}

\affil{$^*$Author to whom any correspondence should be addressed.}

\email{koki.iwasaki@ess.sci.osaka-u.ac.jp}


\begin{abstract}
Accurate measurement of dynamic contact angles in aqueous environments is essential for evaluating surface wettability. However, conventional captive bubble methods often suffer from limitations such as bubble instability and interference from needle wetting. In this study, we develop a pushing–pulling captive bubble method that enables stable and repeatable measurement of dynamic contact angles underwater without directly changing the bubble volume. In this method, a bubble is pushed against and detached from a surface by controlled vertical motion. This procedure allows stable observation of the contact line while suppressing bubble deformation and lateral movement. Dynamic contact angles were measured in both air and water using three types of surfaces: smooth surfaces, sandpaper-polished surfaces prepared to exhibit the Wenzel state in air and the reversed gas–liquid Wenzel state in water, and microstructured surfaces exhibiting hydrophobicity in air. For smooth and Wenzel surfaces, the dynamic contact angles measured in air and water showed similar values. Moreover, the modified captive bubble method exhibited reproducibility comparable to that observed in conventional captive bubble methods under the present experimental conditions. For microstructured surfaces, dynamic contact angle measurements in water had previously been difficult because an air layer remained trapped on the surface. In this study, ultrasonic degassing enabled dynamic contact angle measurements under fully wetted conditions, revealing behavior that differed significantly from that observed in air.
\end{abstract}

\section{Introduction}
Wettability evaluation is one of the simplest and most widely used methods for analyzing surface properties. In the most common approach, a liquid droplet is deposited on a solid surface, and the contact angle is measured\cite{Kwok1999}. The contact angle is defined as the angle between the solid surface and the tangent drawn from the three-phase contact point (gas–liquid–solid) toward the liquid droplet\cite{Kwok1999}. When a droplet advances or recedes on a surface, the corresponding contact angles are referred to as dynamic contact angles, and reliable methods for measuring them have been established\cite{Kwok1999,Furmidge1962}. The difference between the advancing and receding contact angles is called contact angle hysteresis, which plays a critical role in the sliding behavior of droplets on surfaces\cite{Extrand1995}. Contact angle hysteresis arises from local free energy barriers that the contact line must overcome when moving across a surface. As a result, the static contact angle typically lies between the advancing and receding contact angles. Therefore, measurements of dynamic contact angles are useful for evaluating droplet behavior under realistic conditions\cite{McCarthy2007}.

Dynamic contact angles can also be measured in an aqueous environment\cite{Uhlmann2005,Prydatko2018,Grundke2015}. These techniques are generally referred to as the captive bubble method and utilize the attachment of an air bubble to a surface. By employing this method, surface properties can be evaluated even in aqueous environments. In this study, the terms “contact angle in air” and “contact angle in water” refer to the surrounding phase in which the contact angle is measured. Specifically, a contact angle “in air” corresponds to a solid-water-air system in which water is the bounded phase and air is the surrounding unbounded phase, as in a water droplet on a solid surface. In contrast, a contact angle “in water” corresponds to a solid-air-water system in which air is the bounded phase and water is the surrounding unbounded phase, as in a captive air bubble attached to the solid surface. Thus, the two configurations differ in the identity of the bounded phase while sharing the same solid substrate and the same two fluid phases.

Conventional captive bubble methods generally determine advancing and receding contact angles by directly changing the bubble volume through a syringe needle. However, this approach has several limitations. First, wetting between the liquid and the outer wall of the needle can affect the measured contact angle, making it difficult to evaluate the intrinsic wettability of the substrate accurately\cite{Xue2014}. Second, direct expansion and contraction of the bubble often destabilize the bubble shape and can cause lateral motion of the bubble, which introduces errors in contact angle determination and reduces measurement stability\cite{Pogorzelski2013}. In addition, on rough or heterogeneous surfaces, the measured advancing and receding contact angles can depend on bubble size, which makes it difficult to define unique values and further complicates accurate evaluation\cite{Drelich1996}.

To overcome these limitations, a modified captive bubble method was proposed\cite{Pogorzelski2013}. In this method, the bubble is first generated at the syringe tip and then compressed between the syringe and the substrate by moving the syringe vertically, rather than by directly changing the bubble volume through the needle. Because the bubble is mechanically trapped between the syringe tip and the substrate, lateral displacement is effectively suppressed, allowing stable observation of the contact line. In addition, the needle is not used to continuously inject or withdraw air during the measurement, so the influence of needle wetting is eliminated. Since the bubble volume remains essentially constant, shape instability is also reduced. As a result, this pushing-pulling method has the potential to provide more accurate and reproducible measurements of dynamic contact angles in aqueous environments. However, it has not yet been sufficiently verified whether this modified captive bubble method can truly overcome the limitations of conventional methods or whether it provides reliable and reproducible dynamic contact angle measurements under various wetting conditions and surface structures. Therefore, the validity and applicability of this method still require systematic experimental examination.

In this study, dynamic contact angles were measured in both air and water using three types of surfaces: smooth surfaces, sandpaper-polished rough surfaces prepared to clearly exhibit the Wenzel state in both air and water, and microstructured hydrophobic (originally hydrophilic) surfaces that reached the Cassie–Baxter state in air. By comparing the results obtained in these two environments, we demonstrate that the modified captive bubble method is an effective approach for measuring dynamic contact angles underwater.

\begin{figure}[t]
\centering
\begin{subfigure}{0.3\linewidth}
\centering
\includegraphics[height=3.5 cm]{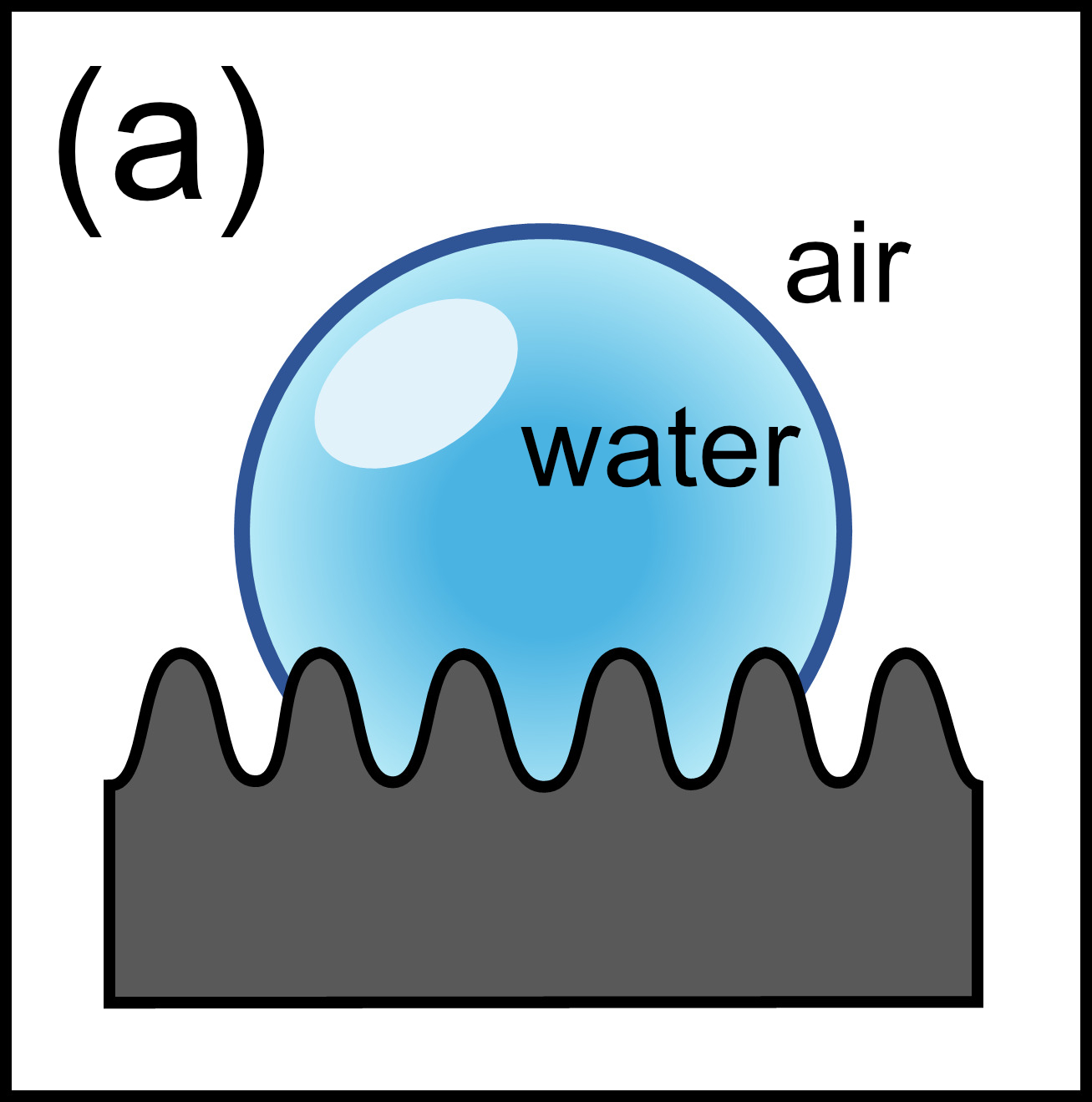}
\end{subfigure}
\hspace{1 mm}
\begin{subfigure}{0.3\linewidth}
\centering
\includegraphics[height=3.5 cm]{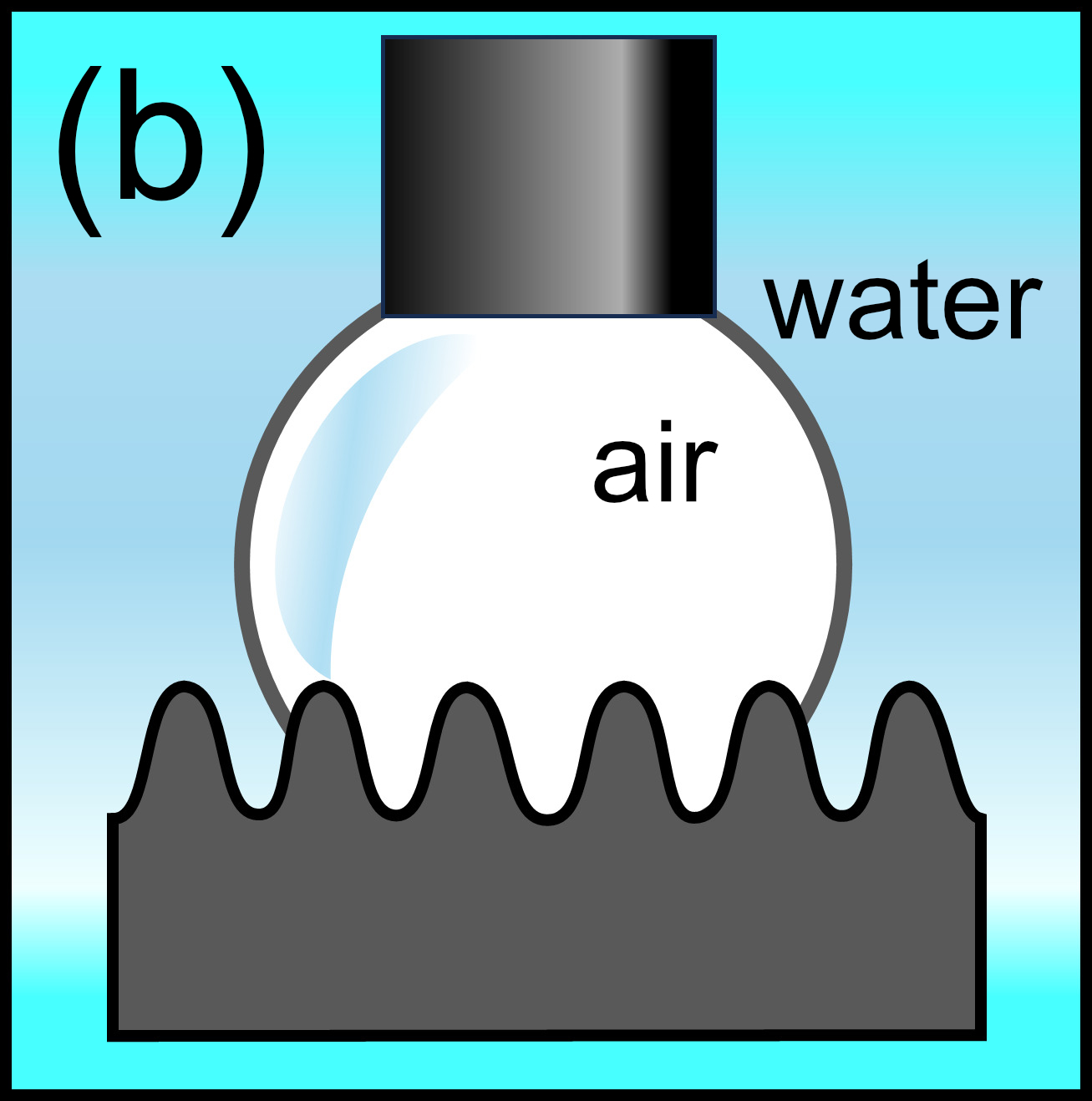}
\end{subfigure}
\hspace{1 mm}
\begin{subfigure}{0.35\linewidth}
\centering
\includegraphics[height=3 cm]{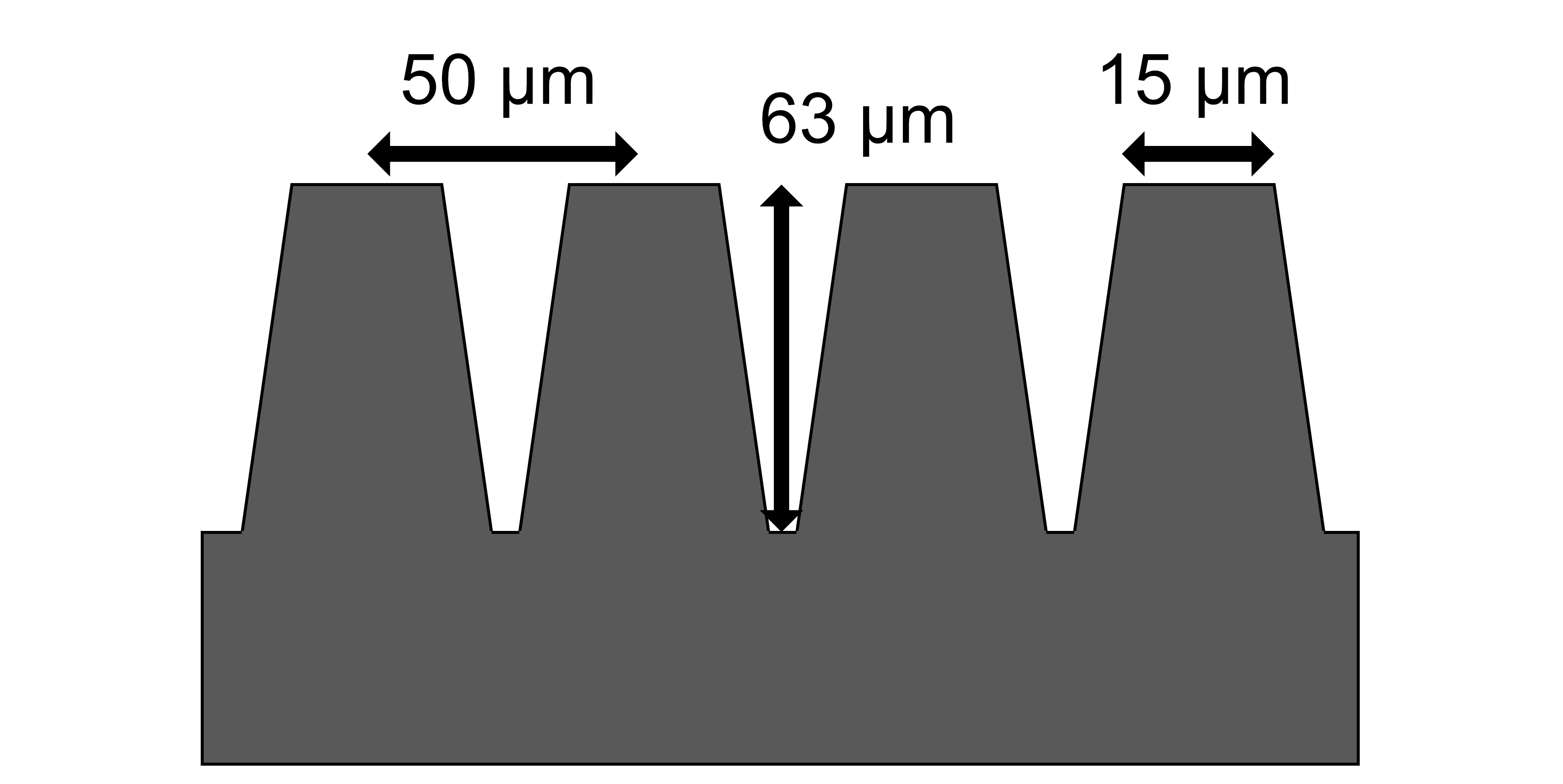}
\end{subfigure}
\caption{Schematic illustrations of the wetting states on a rough surface. (a) Wenzel state in air. (b) Reversed gas–liquid Wenzel state in water. (c)A schematic illustration of the microstructured hydrophobic PMMA surface.}
\label{Wenzel}
\end{figure}

\begin{figure}[b]
\centering
\begin{subfigure}{0.4\linewidth}
\centering
\includegraphics[height=4 cm]{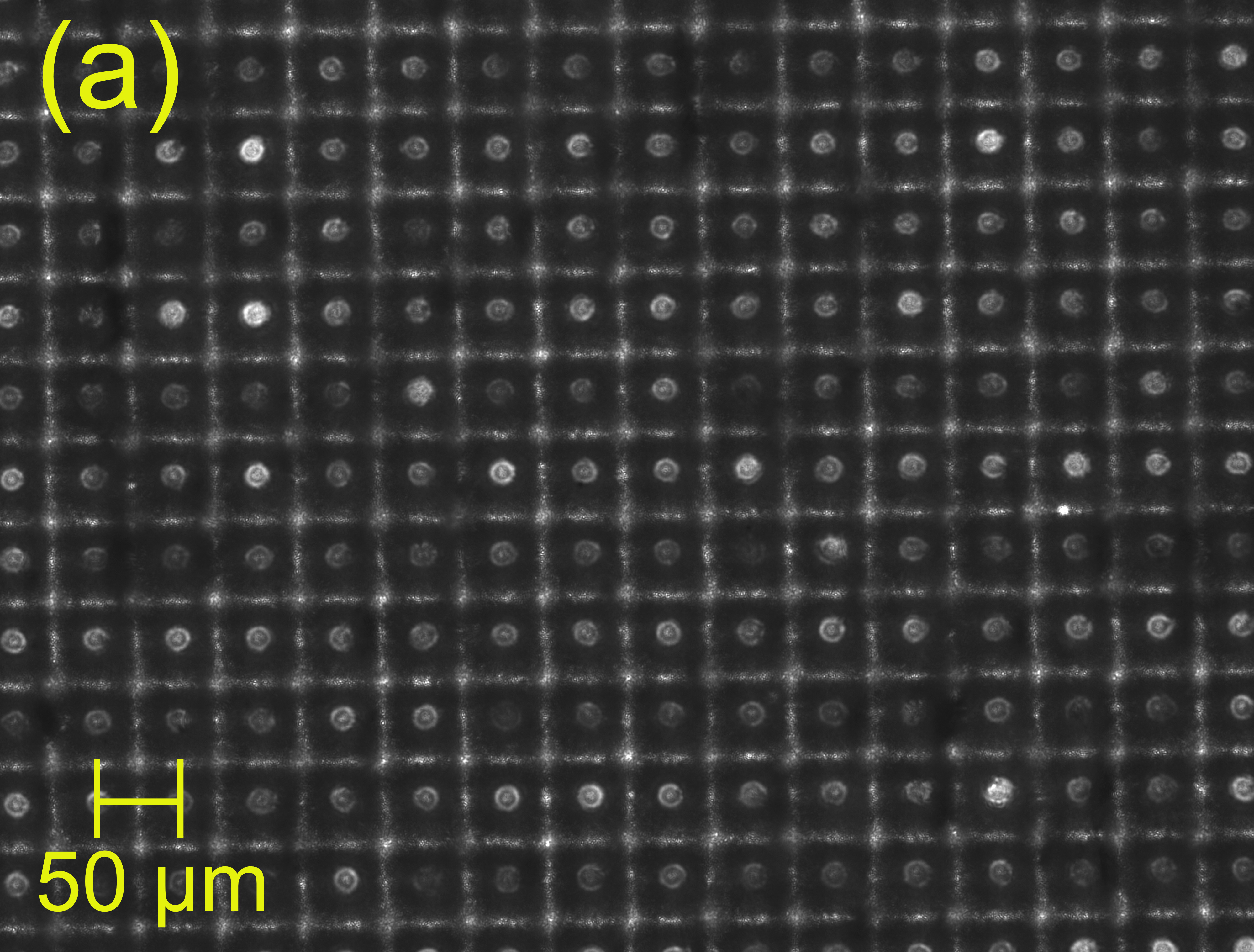}
\end{subfigure}
\hspace{1 mm}
\begin{subfigure}{0.4\linewidth}
\centering
\includegraphics[height=4 cm]{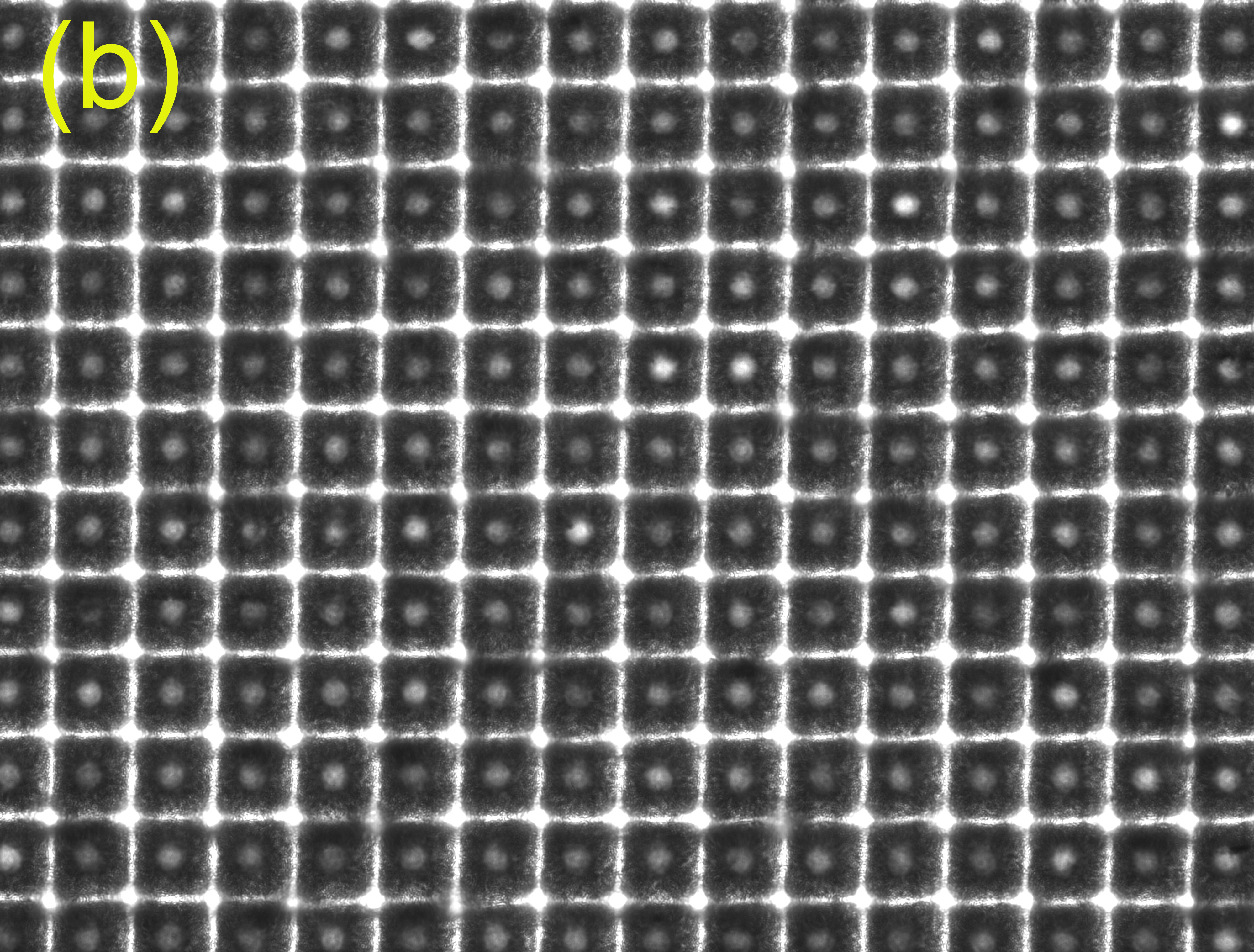}
\end{subfigure}
\caption{Optical microscope(Nikon Ts2R) images of the microstructured PMMA surface (a) before and (b) after ultrasonic treatment in water. Air bubbles trapped in the grooves are observed before the treatment, whereas they disappear after ultrasonic degassing.}
\label{sonic}
\end{figure}

\section{Materials}
The samples measured in this study are smooth plates of PET (polyethylene terephthalate), PC (polycarbonate), PTFE (polytetrafluoroethylene), or PMMA (polymethyl methacrylate). In addition, PET and PMMA plates polished with sandpapers of grit sizes P280, P400, or P600 were prepared. The surface roughness ($R_a$) was measured using a laser displacement sensor (KEYENCE LJ-V7080). The values of $R_a$ of the PET plates polished with P280, P400, and P600 abrasive papers were 8.4, 5.6, and \SI{4.1}{\micro\meter} , respectively. Similarly, the corresponding $R_a$ values for the PMMA plates were 5.8, 3.7, and \SI{3.2}{\micro\meter}, respectively. To examine whether similar dynamic contact angles are obtained in air and water, sandpapers with moderate grit sizes were selected. The resulting rough surfaces were expected to be in the Wenzel state in air and in the reversed gas–liquid Wenzel state in water (Fig. \ref{Wenzel}(a),(b)). PMMA plates with microstructures fabricated by ultrashort-pulse laser processing without coating were also used. The microstructures were fabricated by Hikari Kikai Seisakusyo Co., Ltd. (Mie, Japan). The hydrophobic PMMA plate had regularly arranged pillars on its surface, as shown in Fig. \ref{Wenzel}(c), with a pitch of \SI{50}{\micro\meter} and a depth of approximately \SI{63}{\micro\meter}. The tops of the pillars had a square shape of approximately \SI{15}{\micro\meter} × \SI{15}{\micro\meter}. The surfaces of untreated PET, PTFE, PC, and PMMA plates were regarded as nearly smooth. Prior to the experiments, the smooth PMMA plates and the laser-processed PMMA plates were cleaned by ultrasonic treatment in water. The other samples were cleaned with absolute ethanol, rinsed with water, and then dried. All water used in the experiments was deionized water. The ambient temperature was maintained at \SI{25}{\degreeCelsius}. Although the water temperature was not measured directly, the water was kept under the same temperature-controlled conditions as the ambient environment; therefore, its temperature was assumed to be close to \SI{25}{\degreeCelsius}.

\begin{figure}[t]
\centering
\begin{subfigure}{0.4\linewidth}
\centering
\includegraphics[height=6 cm]{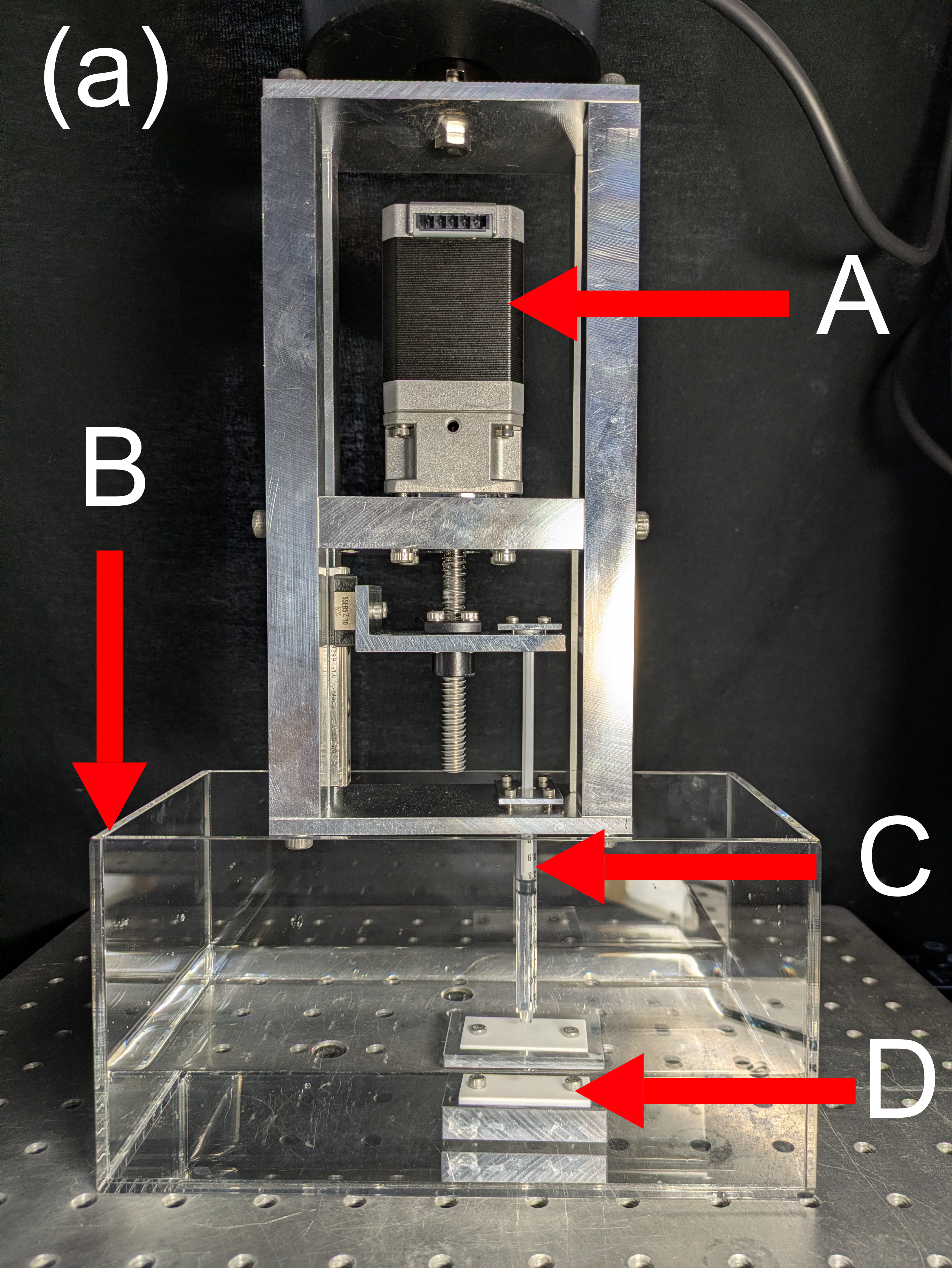}
\end{subfigure}
\hspace{0.1 mm}
\begin{subfigure}{0.4\linewidth}
\centering
\includegraphics[height=6 cm]{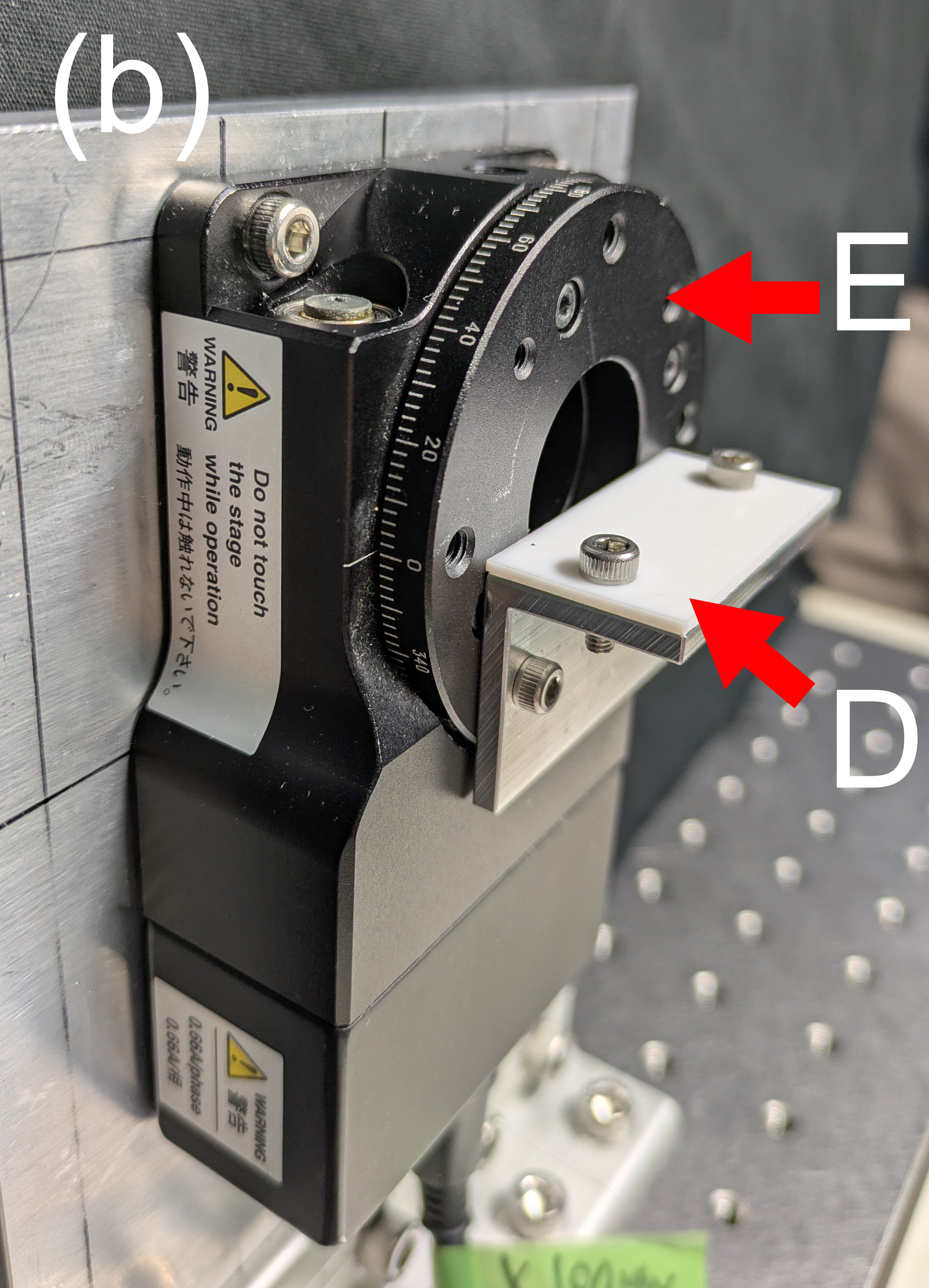}
\end{subfigure}
\caption{Experimental setups used to measure dynamic contact angles. (a) Measurement system for bubbles in water based on the captive bubble method. (b) Measurement system for droplets in air using the tilting stage method. The components indicated by arrows are (A) stepping motor, (B) water tank, (C) syringe, (D) sample, and (E) rotaly stage.}
\label{souchi}
\end{figure}

It is known that when a hydrophobic surface exhibiting the lotus-leaf effect is gently immersed in water, air bubbles may remain trapped in the grooves of the surface structures\cite{Forsberg2011}. However, in practical environments these bubbles are not maintained indefinitely and can easily detach from the grooves when water flows over the surface or after sufficient time has passed\cite{Poetes2010,Wang2014}. In this study, the wettability of hydrophobic surfaces with the lotus-leaf effect was examined in water under conditions where bubbles were not trapped in the grooves. To achieve this, the samples were degassed by ultrasonic treatment after immersion in water. Bright-field optical microscope images (Nikon Ts2R) of the hydrophobic surface before and after ultrasonic treatment are shown in Fig. \ref{sonic}. Before the treatment, the grooves appear dark because air trapped in the structures causes refraction of light. After ultrasonic treatment, the grooves become brighter as the trapped air is removed. These observations indicate that the samples were successfully degassed by the ultrasonic treatment. Although a small amount of gas may remain, the water is sufficiently degassed that its effect can be considered negligible, as described below.

\section{Experiments}

\subsection{Captive bubble method}
We constructed an experimental system to evaluate dynamic contact angles in water using bubble attachment (Fig. \ref{souchi}(a)). First, a water tank was filled with water, and the sample was fixed to a metal plate with screws and immersed in the tank. The entire apparatus was mounted on a universal testing machine (SHIMADZU-AGX), which was used to move the syringe tip into the tank. The plunger was controlled by a stepping motor(Oriental motor PKP246D23A2), which pushed the plunger to generate an air bubble at the syringe tip. The bubble volume in this study was approximately \SI{43}{\micro\liter} (radius $\approx \SI{2.0}{\milli\meter}$). The testing machine was used to position the bubble directly above the target surface. A camera(Nikon D7200) with lens (AI Micro Nikkor 200mm F4S (IF)) was placed so that the bubble could be recorded from the horizontal direction. The image resolution was \SI{3.0}{\micro\meter} per pixel. The total number of pixels was $6000\times4000$. A directional planar light source(OPTEX FA OPF-S51x51W-PS) was placed behind the bubble to illuminate it toward the camera. The syringe was moved downward by 1.3 mm at a speed of 0.5 mm/min to press the bubble against the surface. The syringe was then moved upward at the same speed to detach the bubble from the surface. During the experiments, the behavior of the bubble was recorded every 4~s. The shutter speed of the image acquisition was set to 1/125 s.

\begin{figure}[t]
\centering

\begin{subfigure}{0.3\linewidth}
\centering
\includegraphics[width=4.5 cm]{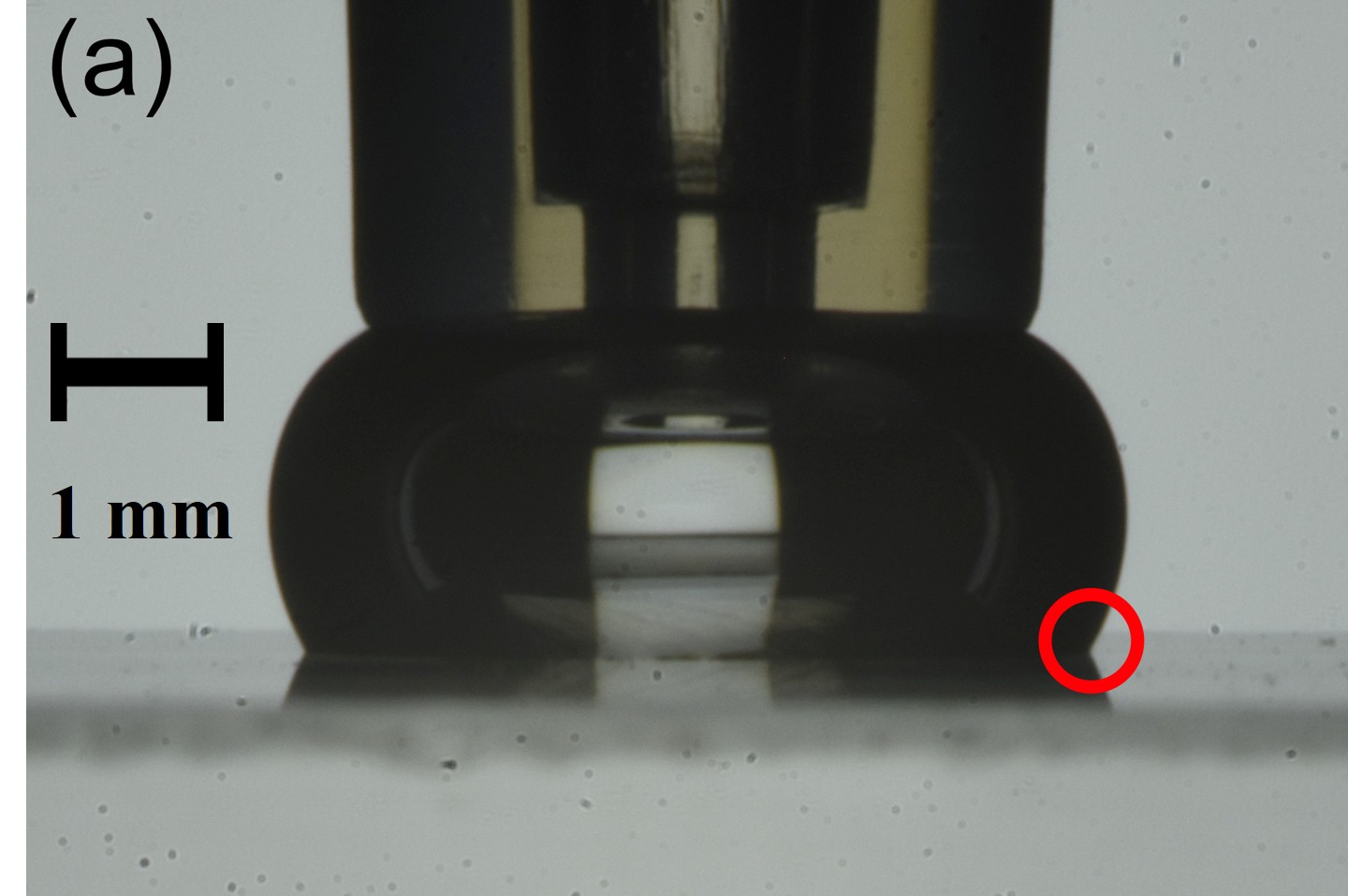}
\end{subfigure}
\hspace{1 mm}
\begin{subfigure}{0.3\linewidth}
\centering
\includegraphics[width=4.5 cm]{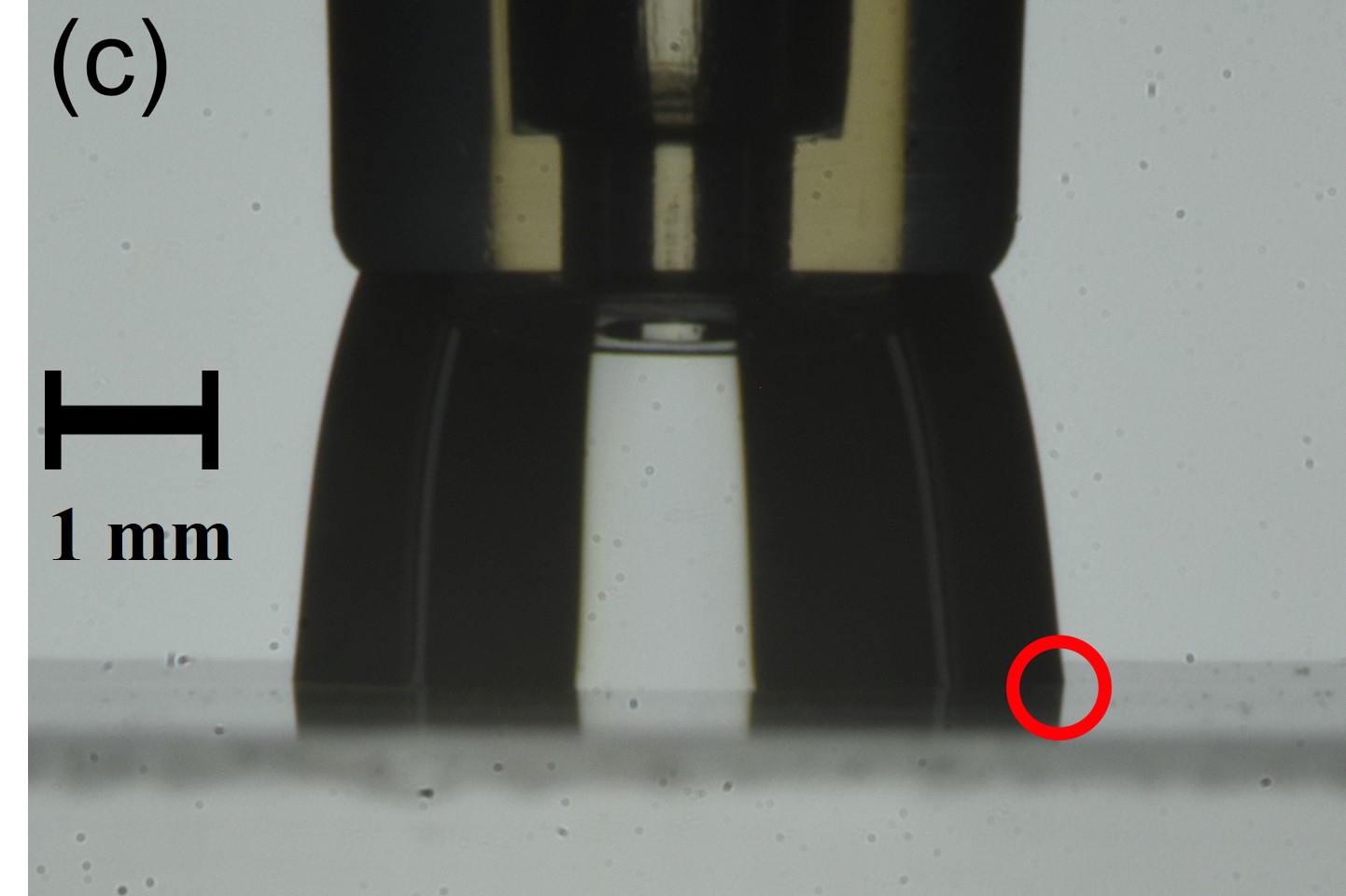}
\end{subfigure}
\hspace{1 mm}
\begin{subfigure}{0.3\linewidth}
\centering
\includegraphics[width=4.5 cm]{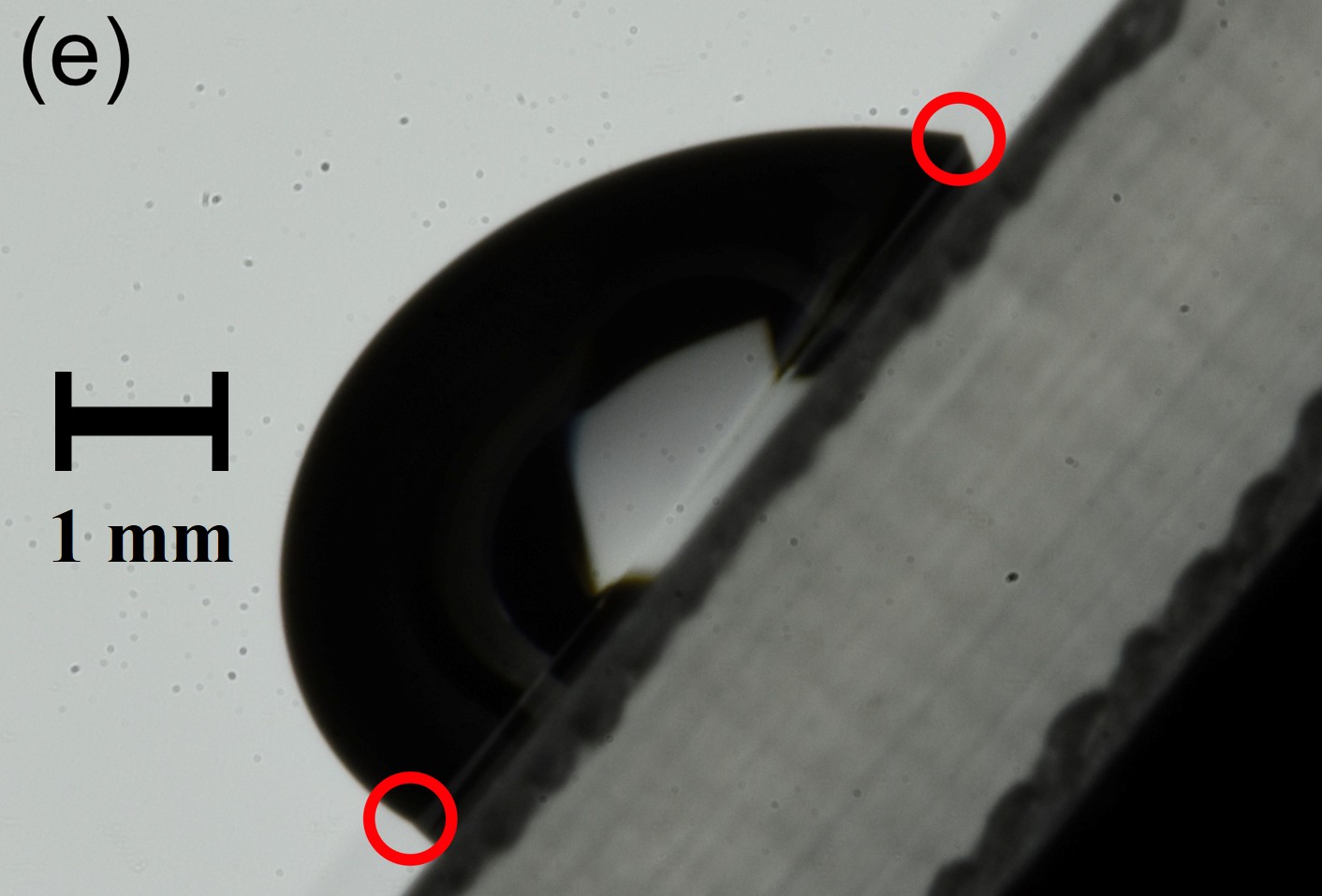}
\end{subfigure}

\vspace{1 mm}

\begin{subfigure}{0.3\linewidth}
\centering
\includegraphics[width=4.5 cm]{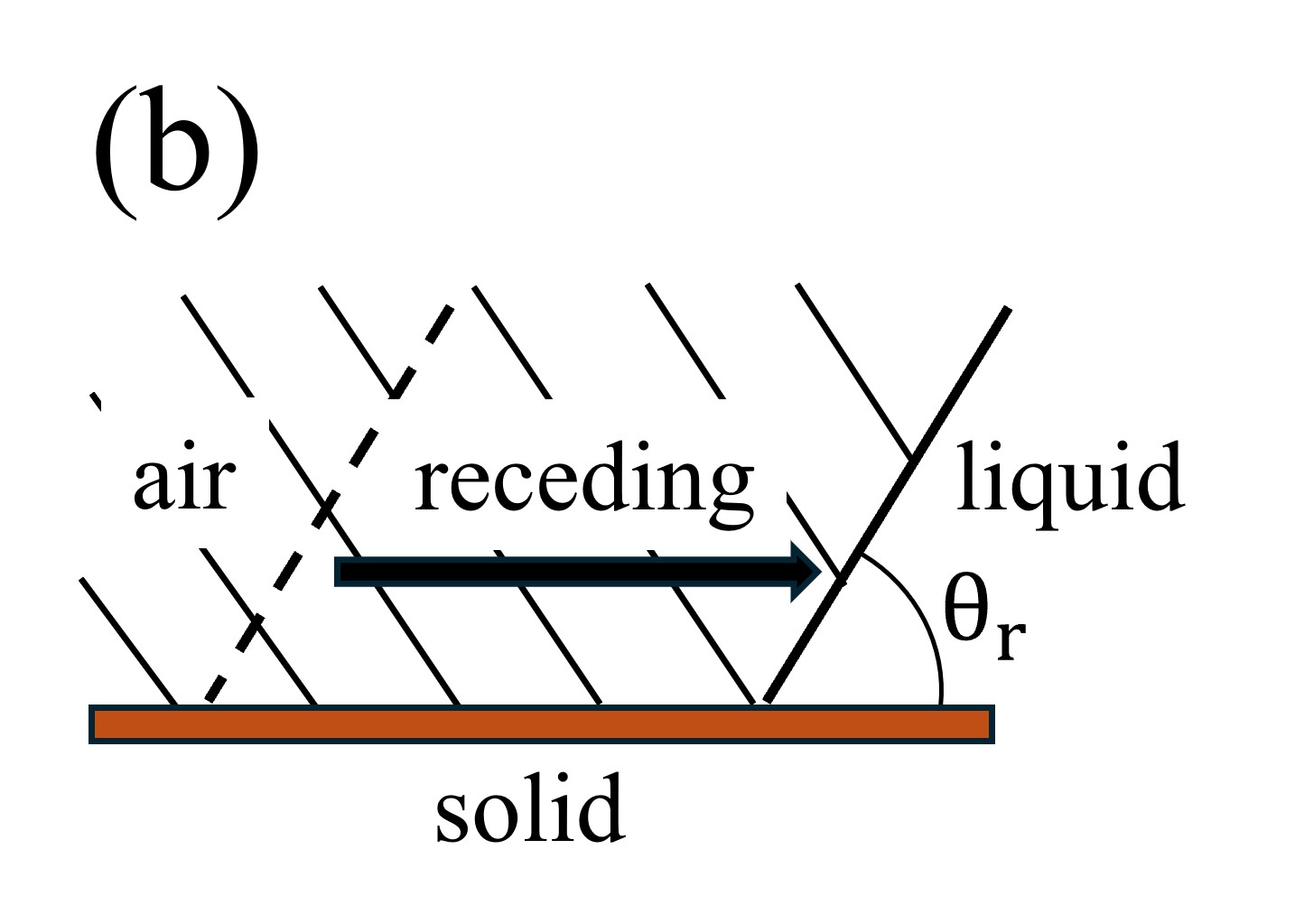}
\end{subfigure}
\hspace{1 mm}
\begin{subfigure}{0.3\linewidth}
\centering
\includegraphics[width=4.5 cm]{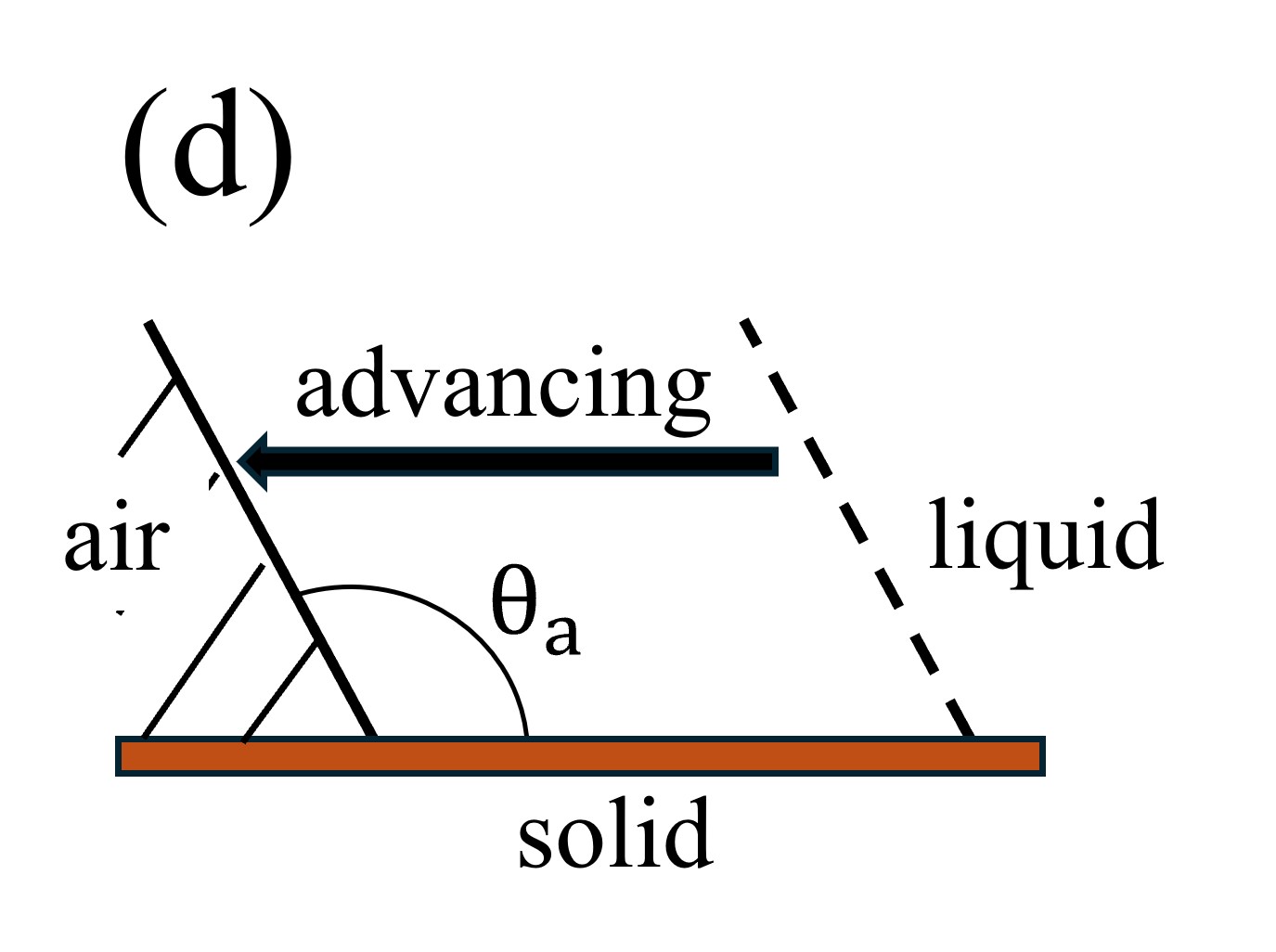}
\end{subfigure}
\hspace{1 mm}
\begin{subfigure}{0.3\linewidth}
\centering
\includegraphics[width=4.5 cm]{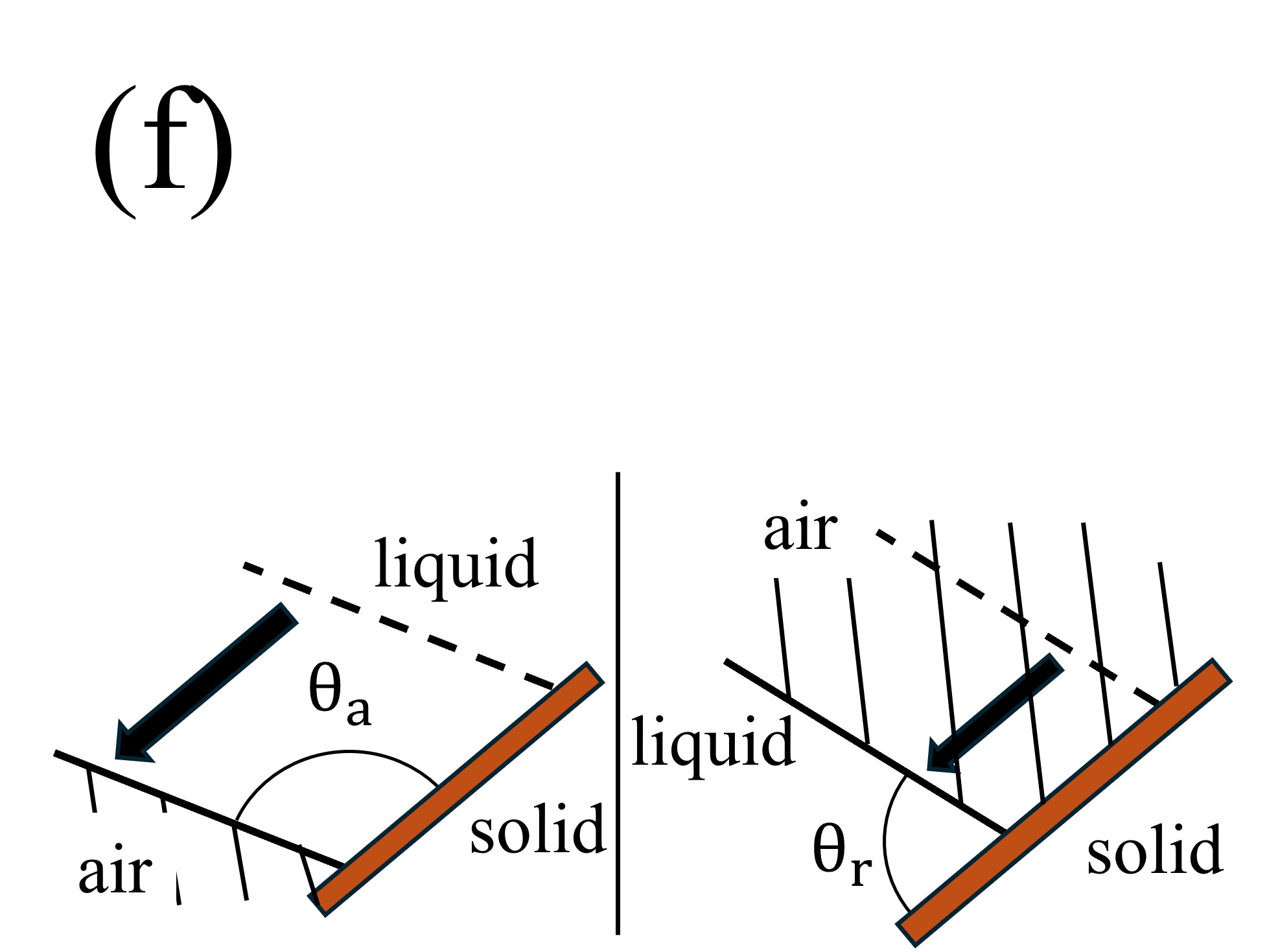}
\end{subfigure}

\caption{Definition of dynamic contact angles. The behavior of the contact line indicated by circles in the image and the corresponding dynamic contact angles are illustrated in the figure below. (a) Pushed bubble exhibiting the receding contact angle. (b) Contact line motion corresponding to the receding contact angle. (c) Pulled bubble exhibiting the advancing contact angle. (d) Contact line motion corresponding to the advancing contact angle. (e) Sessile droplet on a tilted substrate exhibiting the advancing and receding contact angles. (f) Contact line motion corresponding to the advancing and receding contact angle.}
\label{definition}

\end{figure}

The capillary number $C_A=\mu U/\gamma$ is estimated $1.1\times10^{-6}$. Here, $\mu$ is the viscosity of water ($1.00\times 10^{-3}$ Pa$\cdot$s), $U$ is the characteristic velocity ($U=0.5$ mm/min $=8.33\times 10^{-5}$ m/s), and $\gamma$ is the water–air interfacial tension ($7.28\times 10^{-2}$ N/m). This small $C_A$ indicates that the experiments were performed under sufficiently quasi-static conditions. The Bond number $B_O=\Delta \rho gL^2/\gamma$ is estimated as 2.2. Here, $\Delta \rho$ is the density difference between water and air ($9.98\times 10^2$ $\mathrm{kg/m^3}$), $g$ is the gravitational acceleration ($9.81$ $\mathrm{m/s^2}$), and $L$ is the characteristic length (the bubble diameter of approximately 4 mm). This large $B_O$ indicates that gravity was somewhat dominant. However, according to previous research\cite{Drelich1996}, when the bubble size is too small, its shape can become unstable due to the influence of surface roughness, and the bubble can be strongly affected by local surface conditions. These effects become particularly pronounced for bubbles with a radius of less than 1 mm\cite{Drelich1996}. Therefore, in the present study, we used bubbles with a sufficiently large radius of approximately 2 mm to minimize the influence of local surface conditions and obtain stable bubble shapes. In addition, the contact angle is determined at a length scale of several tens of micrometers, where capillary forces are dominant. Indeed, if $L=0.1$ mm is used, $B_O=1.4\times10^{-3}$ is obtained. Therefore, when considering the region in the vicinity of the contact line, it is conventional to neglect the effect of gravity.

The advancing and receding contact angles are defined based on whether the liquid phase advances or recedes relative to the gas phase. Fig. \ref{definition}(a) shows the image of the bubble being pushed against the substrate. The motion of the contact line at the location indicated by the red circle is schematically shown in Fig. \ref{definition}(b). Although the bubble advances toward the liquid phase, the liquid phase recedes relative to the gas phase. Therefore, the contact angle measured under this condition is defined as the receding contact angle. Conversely, Fig. \ref{definition}(c) shows the image of the bubble being pulled away from the substrate. The motion of the contact line at the location indicated by the red circle is shown in Fig. \ref{definition}(d). In this case, the liquid phase advances relative to the gas phase, and thus the measured contact angle is defined as the advancing contact angle.

\subsection{Droplet tilting method}
We also constructed an experimental setup for measuring dynamic contact angles in air based on droplet tilting method (Fig. \ref{souchi}(b))\cite{Furmidge1962}. A rotation stage(COMS PS60BB-360R) with a transmission hole at its center was mounted vertically and controlled by a PC. A metal plate for fixing the sample was attached to the rotation stage, and the sample was secured with screws. A camera was placed in front of the stage so that the droplet could be recorded horizontally. The identical planar light source behind the stage illuminated the droplet through the transmission hole toward the camera. A water droplet(approximately \SI{33}{\micro\liter}) was deposited on the sample surface, and the stage was tilted at a rate of \SI{0.25}{\degree\per\second} until the droplet slid off. The droplet behavior was recorded once per second during the experiment. The droplet at the moment of sliding is shown in Fig. \ref{definition}(e). The contact angle at the upper side of the droplet, where the contact line is receding, is defined as the receding contact angle, whereas that at the lower side, where the contact line is advancing, is defined as the advancing contact angle. The corresponding contact line motion is schematically shown in Fig. \ref{definition}(f). Each experiment was repeated three times, and the average values were used.

\begin{figure}[t]
\centering

\begin{subfigure}{0.4\linewidth}
\centering
\includegraphics[width=5.5 cm]{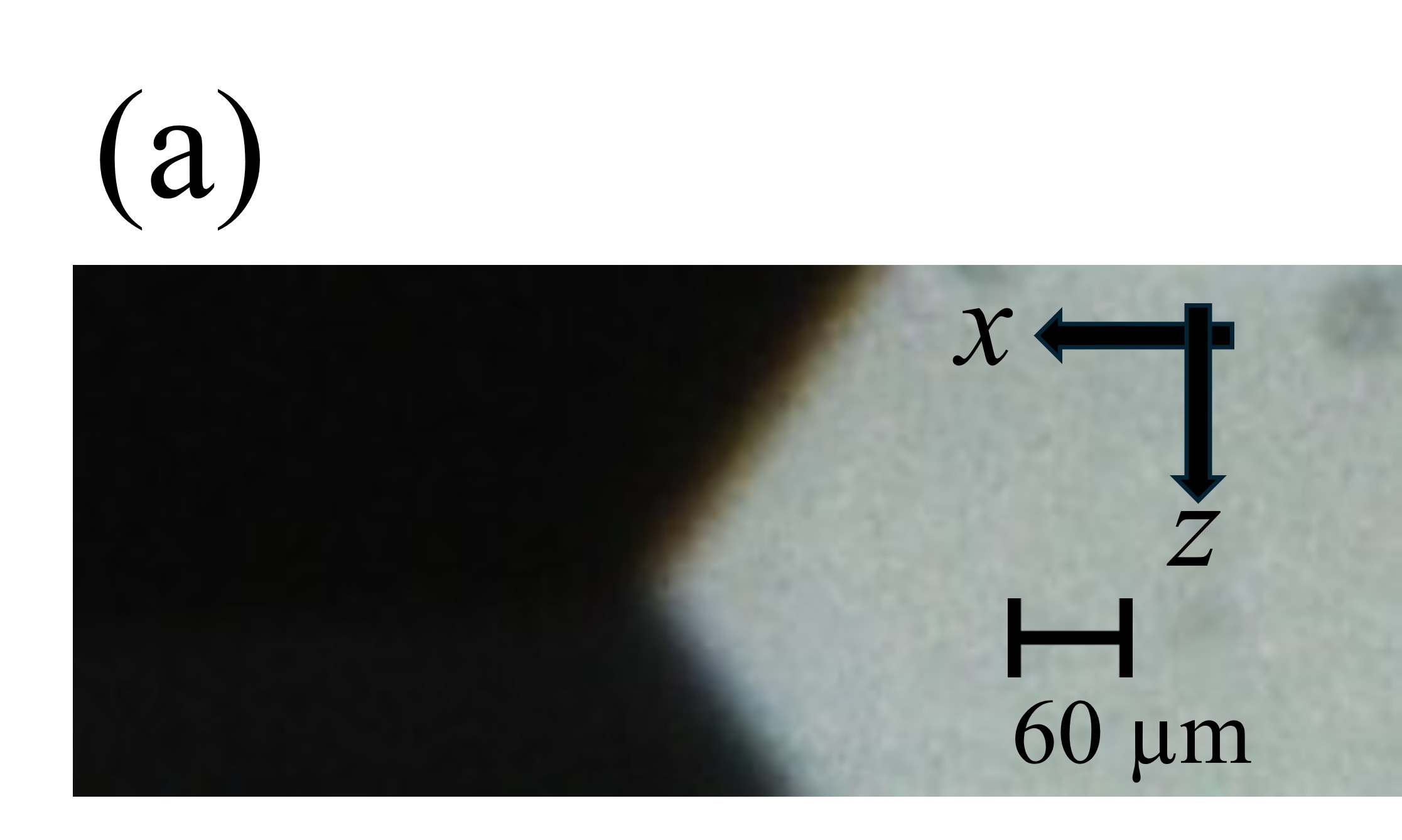}
\end{subfigure}
\hspace{1 mm}
\begin{subfigure}{0.5\linewidth}
\centering
\includegraphics[width=7.5 cm]{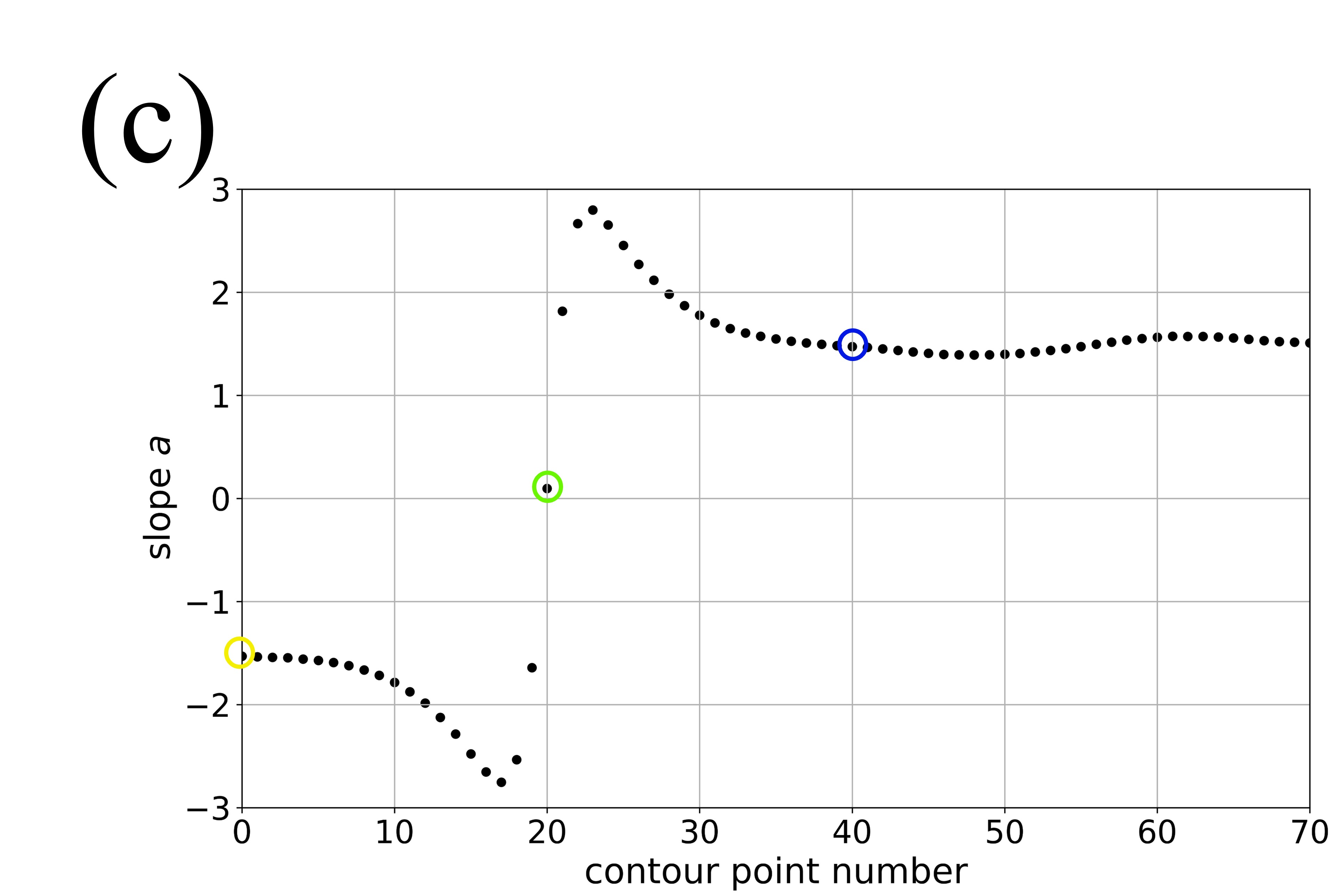}
\end{subfigure}
\vspace{1 mm}
\begin{subfigure}{0.45\linewidth}
\centering
\includegraphics[width=5.5 cm]{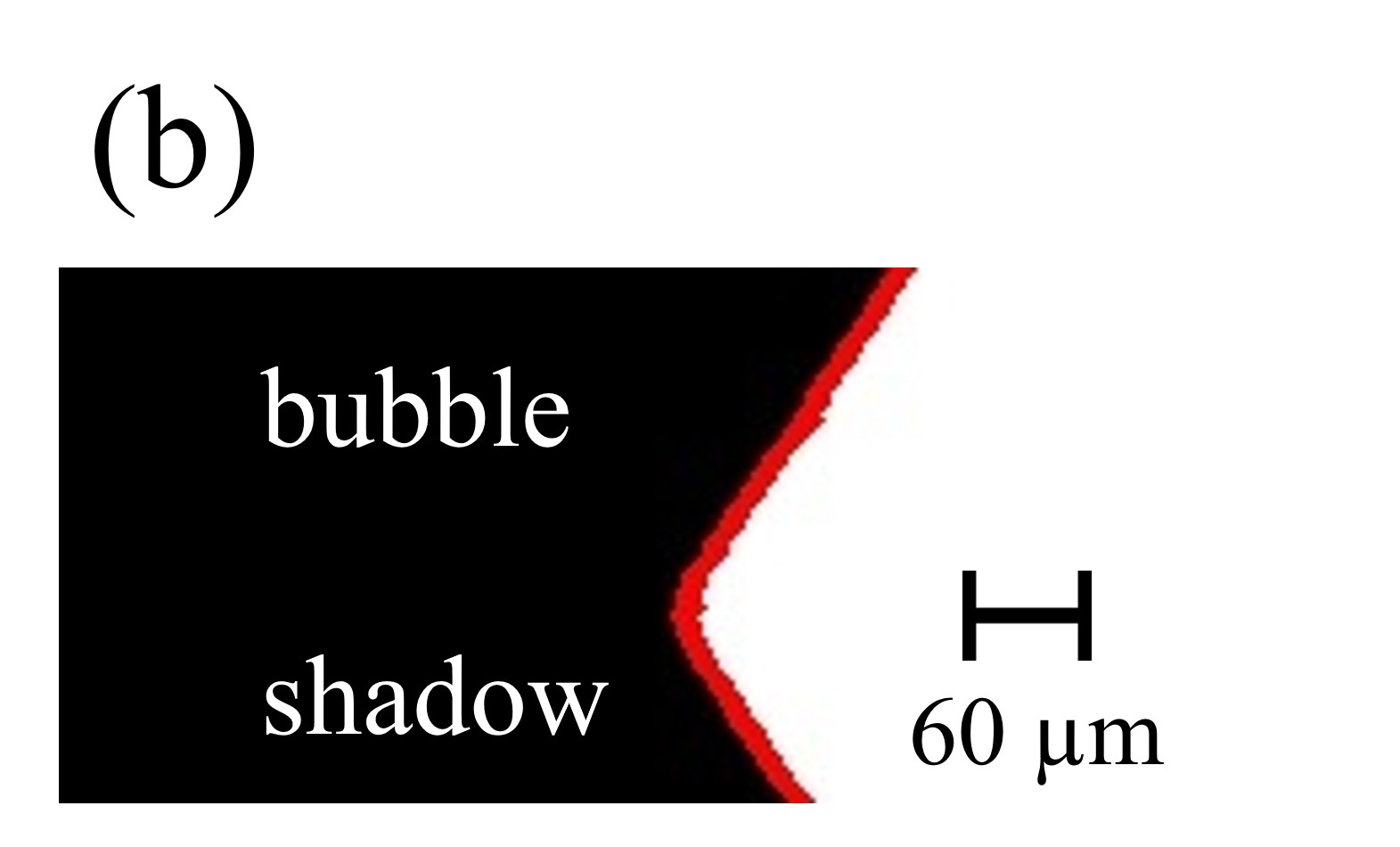}
\end{subfigure}
\hspace{1 mm}
\begin{subfigure}{0.45\linewidth}
\centering
\includegraphics[width=5.5 cm]{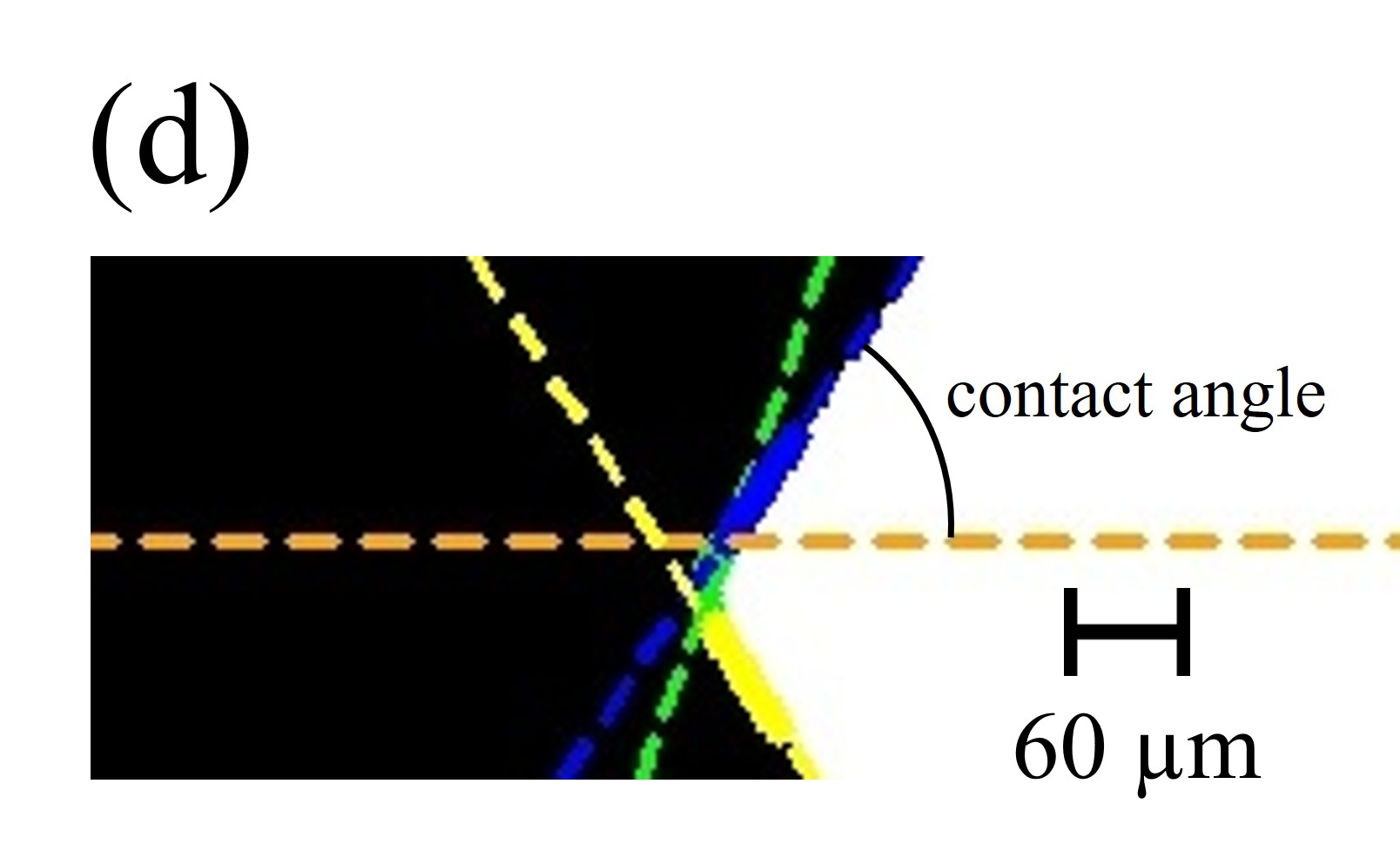}
\end{subfigure}
\hspace{1 mm}
\caption{(a) A region of interest (ROI) containing the contact region between the bubble (or droplet) and the substrate (ROI size: $250\times100$ pixels). (b) Binarized ROI image with the contour points plotted in red. (c) Slope $a$ of the fitting line corresponding to each contour point is plotted. (d) The fitting lines obtained from three sets of 20 contour points: around the slope sign-change position (green), shifted 20 points in the positive $z$-direction from the sign-change position (yellow), and shifted 20 points in the negative $z$-direction from the sign-change position (blue). The blue fitting line was selected as it uses points closest to the contact line while minimizing the inclusion of boundary points. The angle between the blue fitting line and the horizontal orange line was measured as the contact angle.}
\label{analysis}

\end{figure}

\subsection{Analysis method}
In the present experiments, the contact angle on the right side of the bubble (or droplet) was measured. First, a region of interest (ROI: $250\times100$ pixels) containing the contact region between the bubble and the substrate was extracted from the original image (Fig. \ref{analysis}(a)). The ROI was then converted to 8 bit grayscale using OpenCV and binarized using a threshold value of 90. As a result, the bubble (or droplet) and its shadow were represented by black pixels, while the remaining regions were represented by white pixels. The edge was detected by identifying the first black pixel encountered when scanning the image from the right edge, and the contour was extracted accordingly (red curve in Fig. \ref{analysis}(b)). The contour points were then ordered according to their $z$-coordinates, starting from the point with the largest z-coordinate, where the $z$-axis was defined as positive downward and $x$ axis compounds to horizontal direction (Fig. \ref{analysis}(a)). The contour was thus treated as a continuous sequence of points. Local linear regression was performed for each set of 20 consecutive contour points to obtain the form $z=ax+b$, where $a$ and $b$ are fitting parameters, and $x$ denotes the holizontal axis (Fig. \ref{analysis}(a)). Starting from the point with the largest $z$-coordinate, the the fitted line was calculated. This procedure was repeated for all contour points, thereby assigning a local slope $a$ to each consecutive contour point. The reordered contour points were numbered sequentially from 0, and the slope of the fitting line corresponding to each point is shown in Fig. \ref{analysis}(c). The slope diverges near the boundary between the bubble (or droplet) and its shadow, with its sign changing across this region. Fig. \ref{analysis}(d) shows the fitting lines obtained using three sets of contour points: 20 contour points around the position where the sign of the slope of the fitting line changes (green), 20 contour points shifted by 20 points in the positive $z$-direction from the sign-change position (yellow), and 20 contour points shifted by 20 points in the negative $z$-direction (blue). The positions in Fig. \ref{analysis}(c) corresponding to each fitting line are indicated in the same color as the respective lines. To perform the linear fitting using contour points as close as possible to the contact line while minimizing the inclusion of points near the boundary, the blue fitting line was selected. At the points indicated in blue, the influence of the boundary is largely eliminated, and the slope is observed to have stabilized. Therefore, for all datasets, the fitting line at a point shifted 20 points in the negative $z$-direction from the slope sign-change position was selected. The angle between this fitting line and the horizontal line shown in orange was then measured as the contact angle.

\begin{figure}[t]
\centering

\begin{subfigure}{0.3\linewidth}
\centering
\includegraphics[width=4.5 cm]{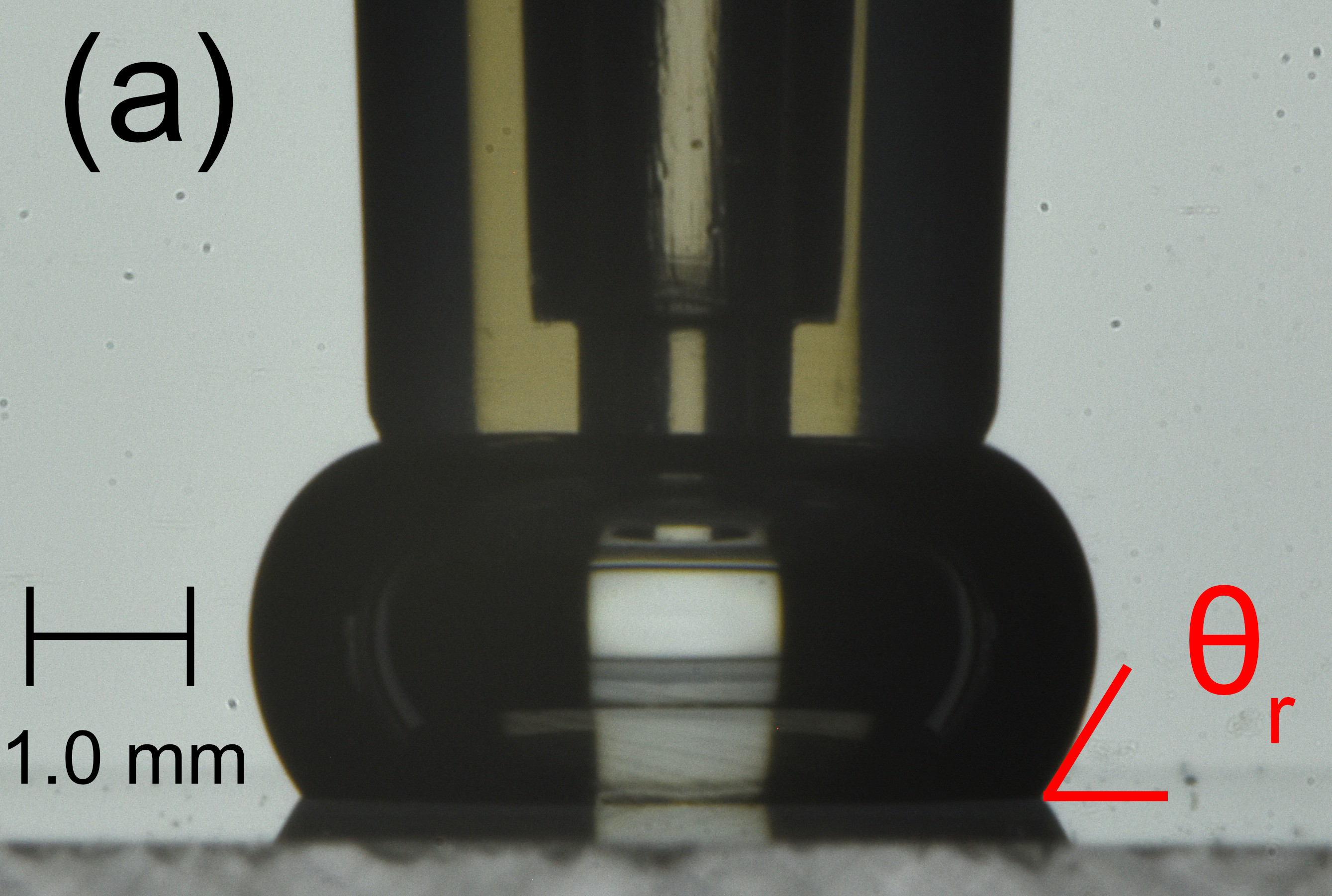}
\end{subfigure}
\hspace{1 mm}
\begin{subfigure}{0.3\linewidth}
\centering
\includegraphics[width=4.5 cm]{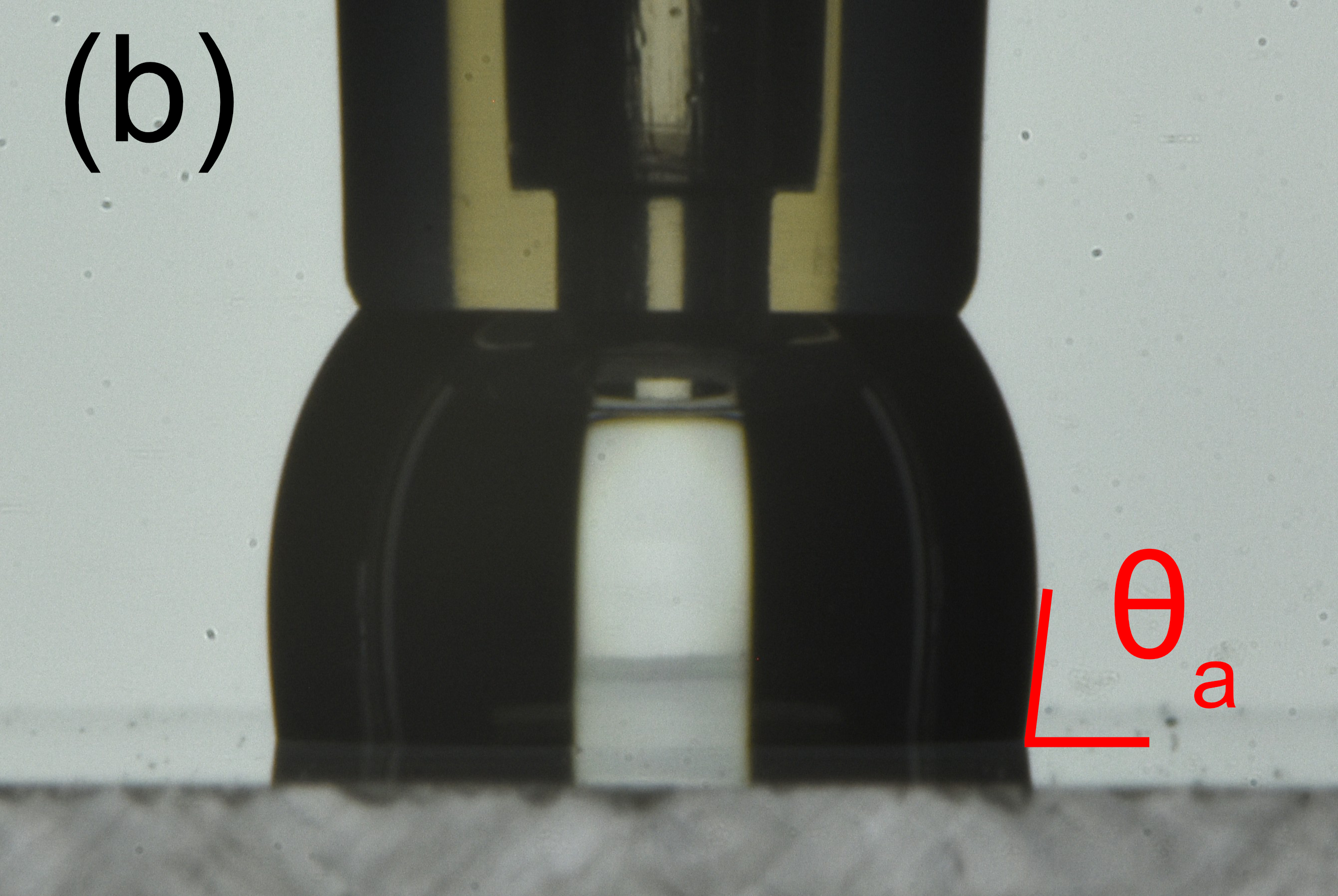}
\end{subfigure}
\hspace{1 mm}
\begin{subfigure}{0.3\linewidth}
\centering
\includegraphics[width=4.5 cm]{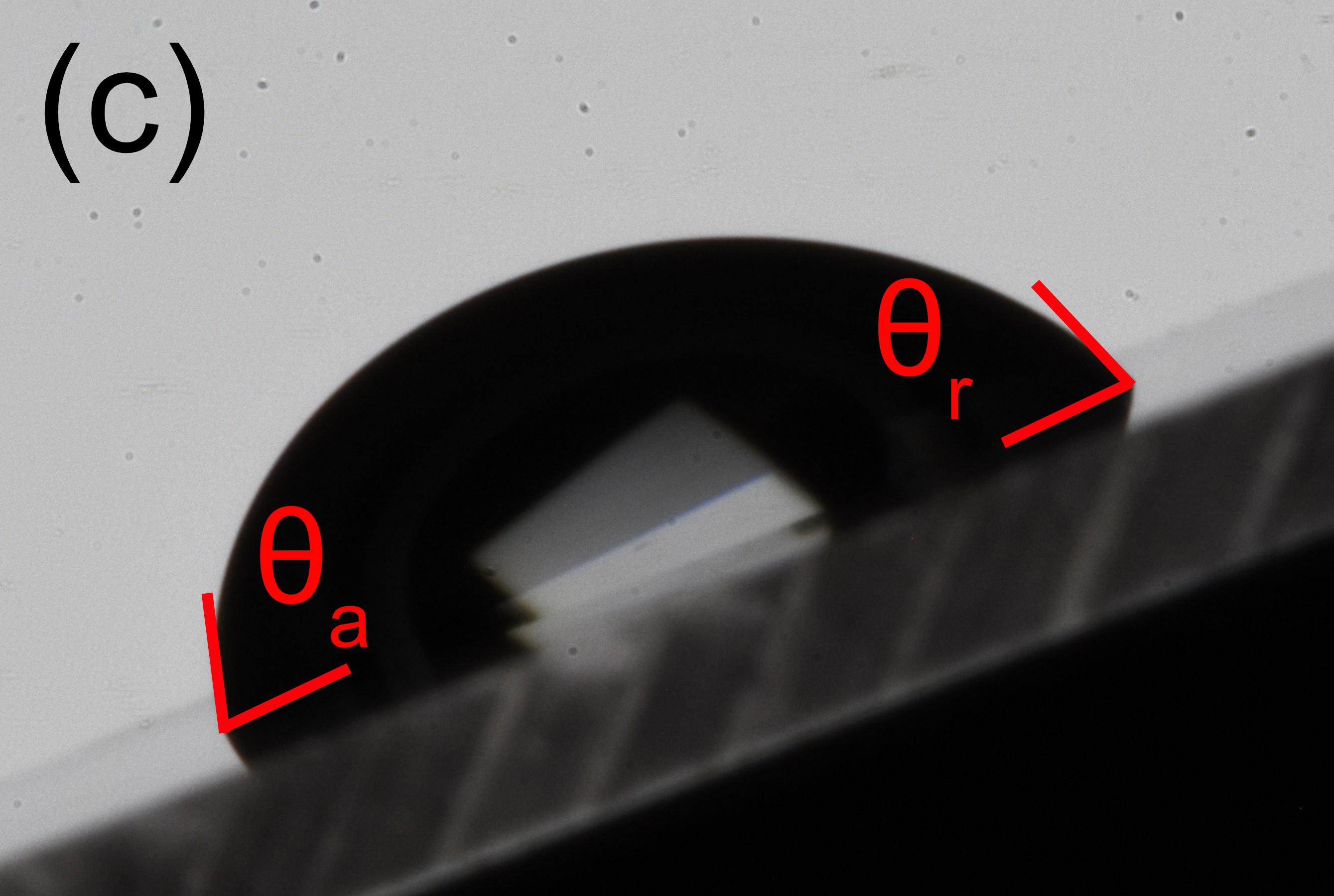}
\end{subfigure}

\vspace{1 mm}

\begin{subfigure}{1.0\linewidth}
\centering
\includegraphics[height=6 cm]{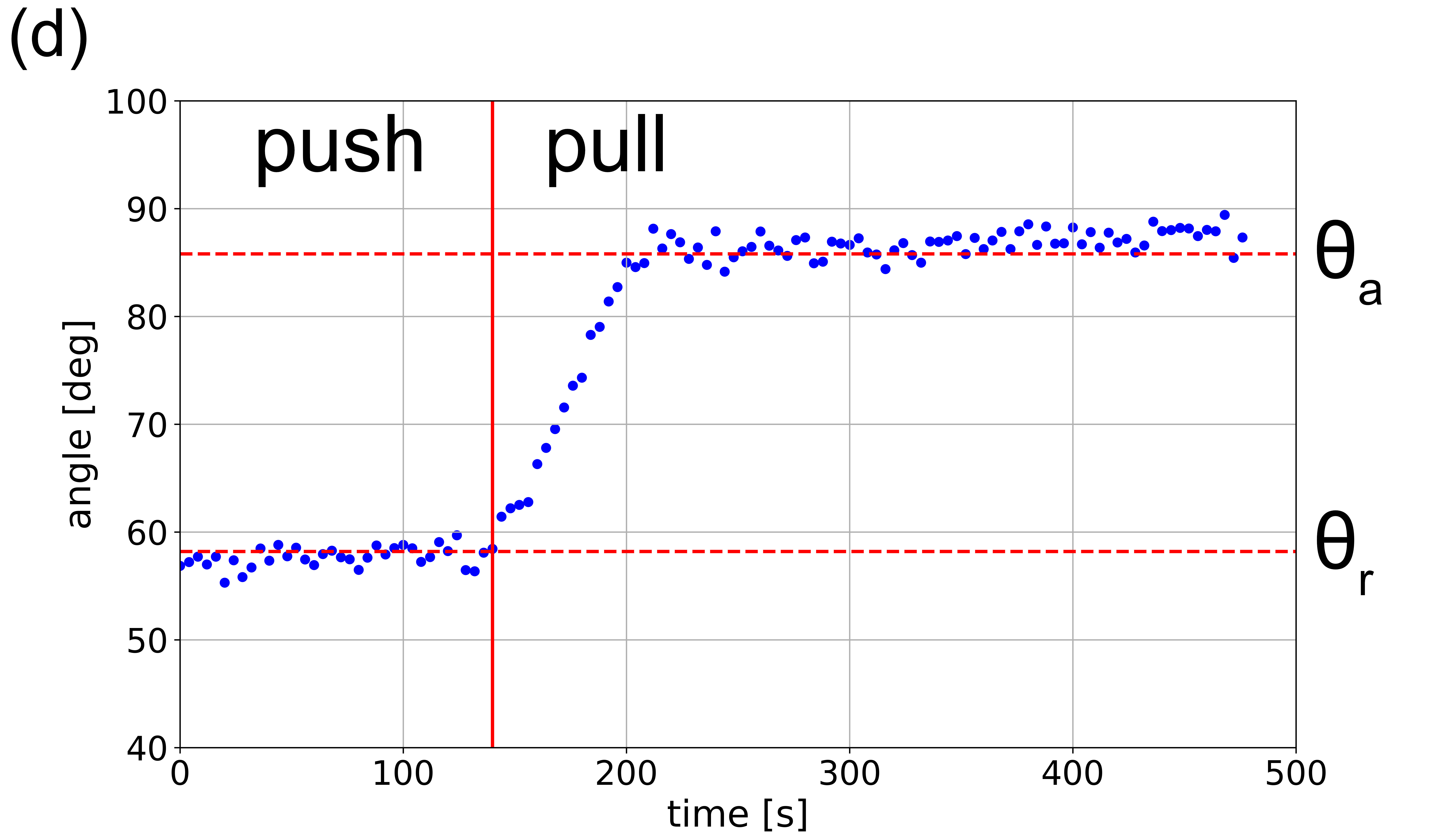}
\end{subfigure}

\caption{Representative measured data of dynamic contact angles on a PET surface. (a) Receding contact angle (\textit{$\theta_r$}) of a bubble during pushing against the surface. (b) Advancing contact angle (\textit{$\theta_a$}) of a bubble during pulling away from the surface. (c) Advancing and receding contact angles (\textit{$\theta_a$} and \textit{$\theta_r$}) of a water droplet measured in air just before sliding. (d) Time variation of the dynamic contact angle of bubble during the pushing and pulling processes of a bubble on a PET surface.}
\label{PETresults}

\end{figure}

The edge-detection error was evaluated by varying the binarization threshold by 90±10. The contact line position identification error was evaluated by varying the number of data points between the slope sign-change position and the fitting region by 20±10 (10 pixels: approximately \SI{30}{\micro\meter}). The fitting error was determined from the standard error of the slope obtained by linear regression using the least-squares method and was converted to the contact-angle error using $\theta =\mathrm{tan}^{-1}(a)$. The propagated error was calculated as the square root of the sum of the squares of these individual errors.

\begin{figure}[t]
\centering

\begin{subfigure}{0.3\linewidth}
\centering
\includegraphics[height=6 cm]{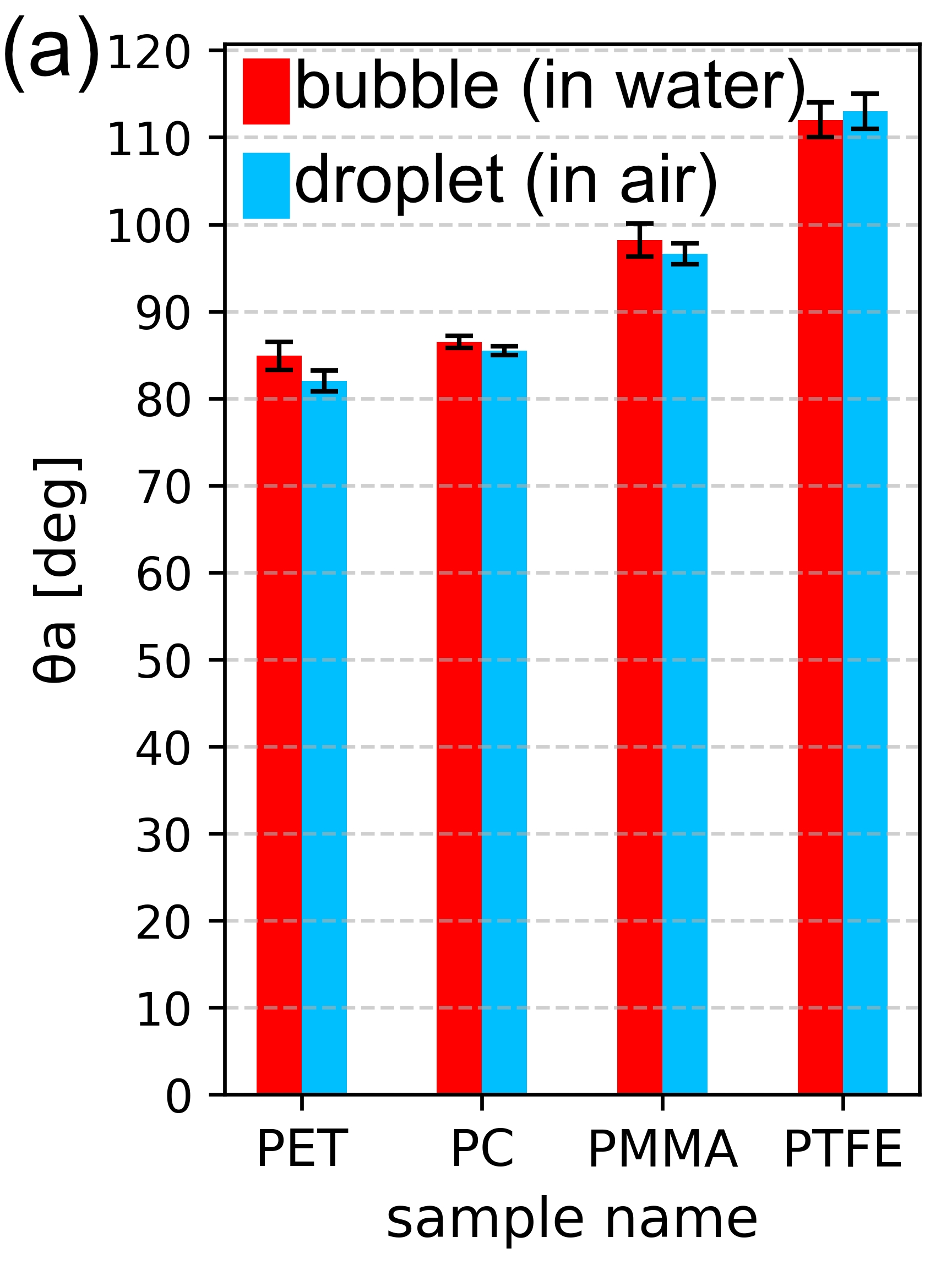}
\end{subfigure}
\hspace{1 mm}
\begin{subfigure}{0.3\linewidth}
\centering
\includegraphics[height=6 cm]{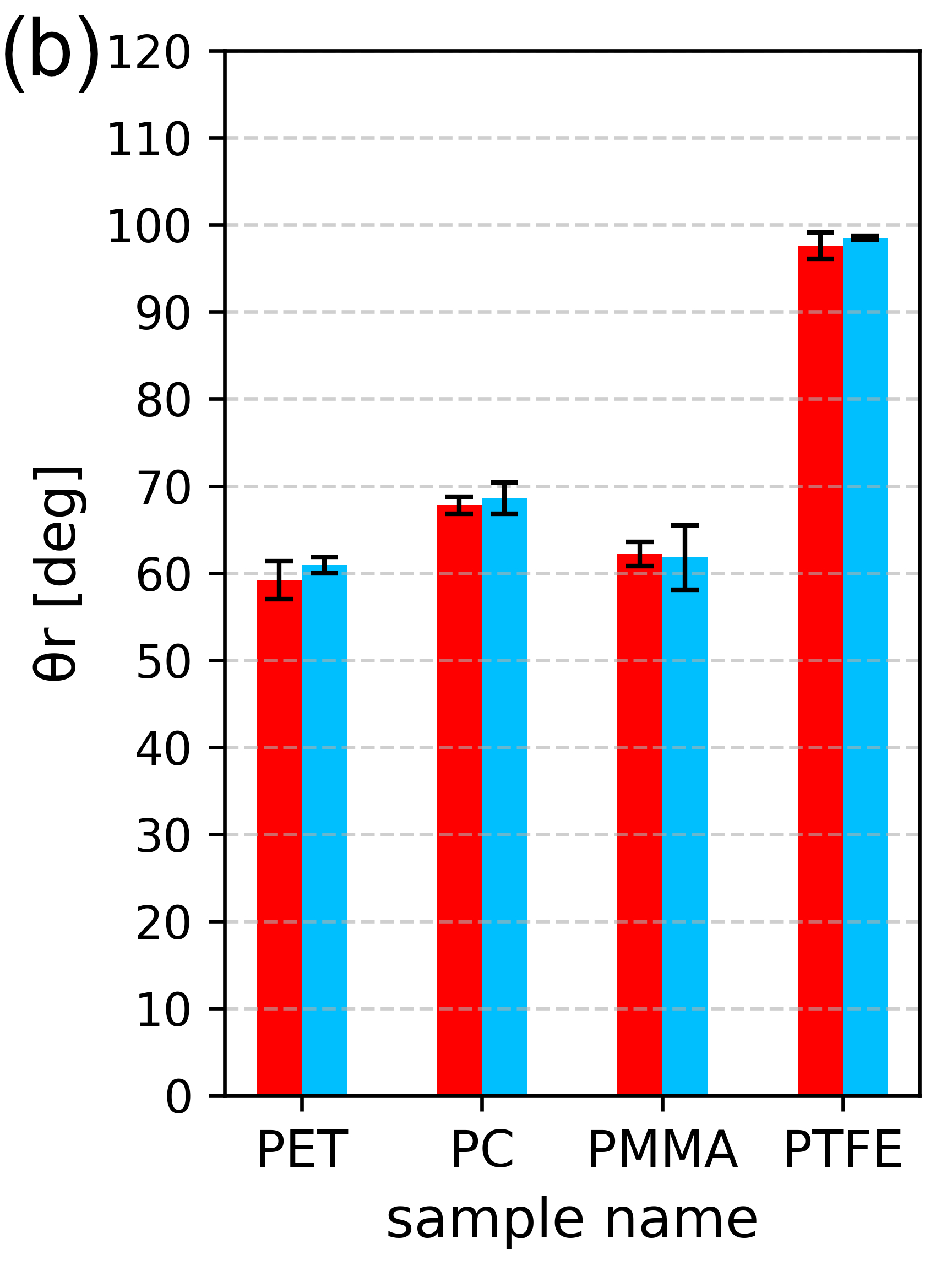}
\end{subfigure}
\hspace{1 mm}
\begin{subfigure}{0.3\linewidth}
\centering
\includegraphics[height=6 cm]{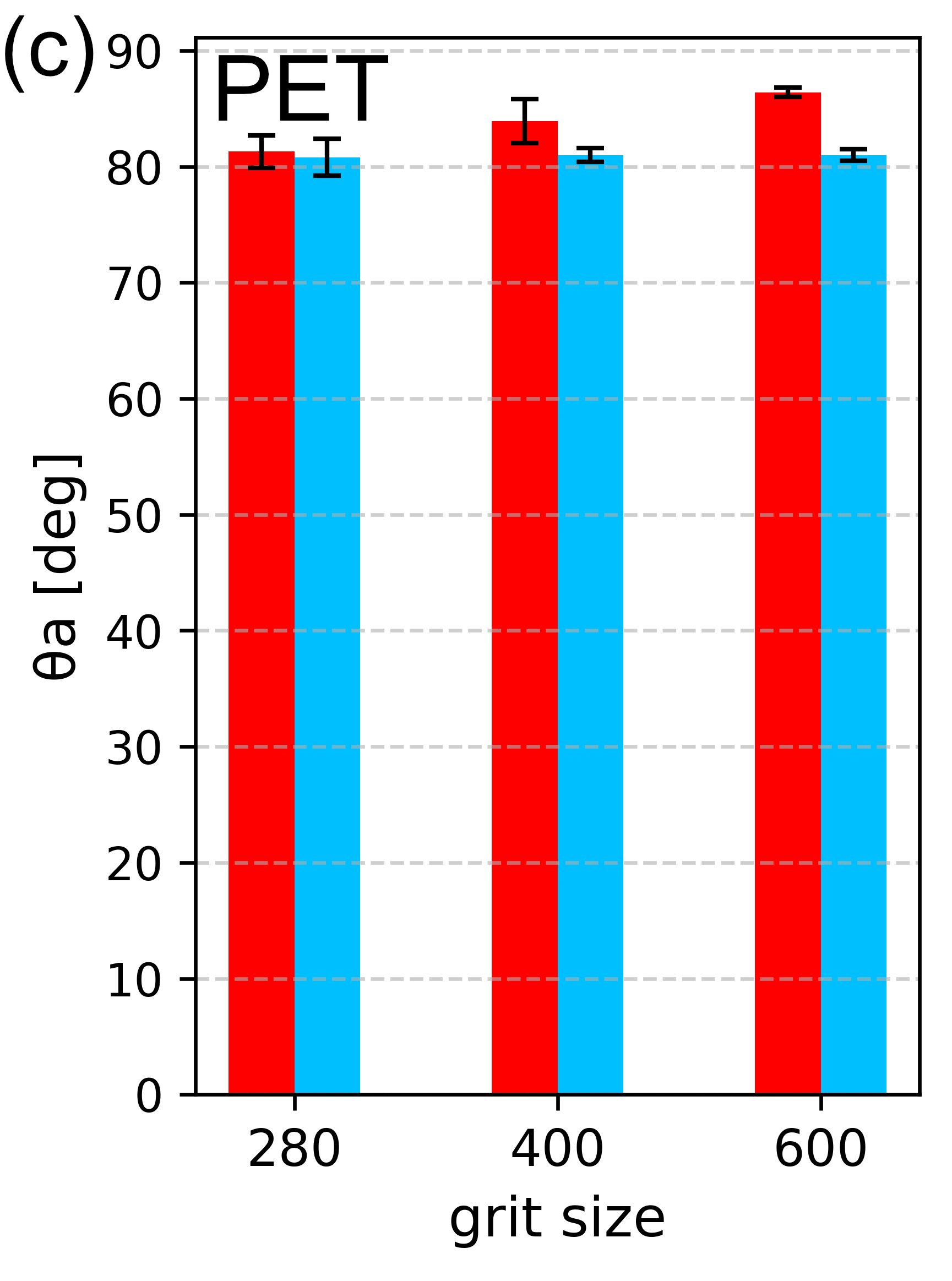}
\end{subfigure}

\vspace{1 mm}

\begin{subfigure}{0.3\linewidth}
\centering
\includegraphics[height=6 cm]{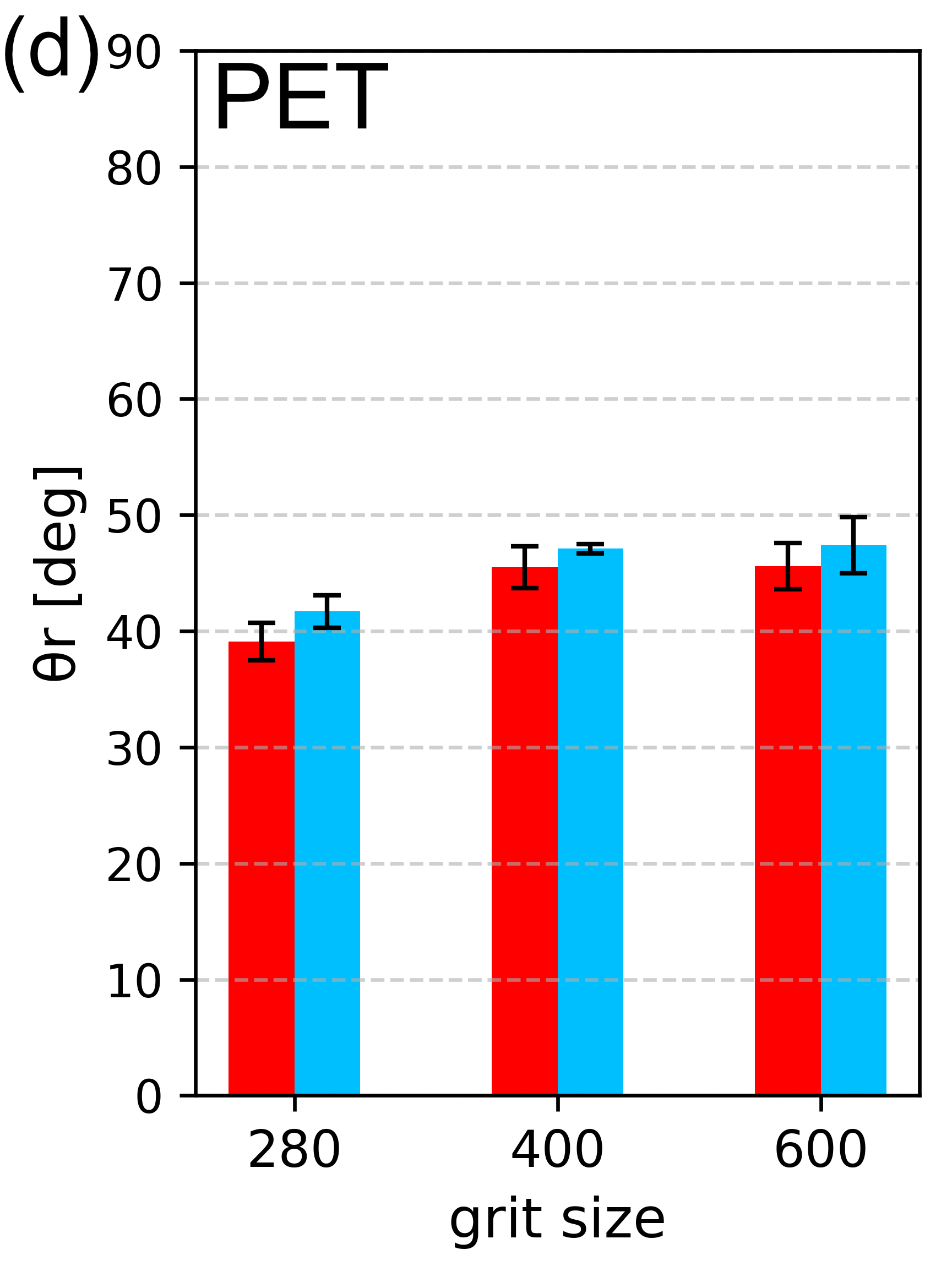}
\end{subfigure}
\hspace{1 mm}
\begin{subfigure}{0.3\linewidth}
\centering
\includegraphics[height=6 cm]{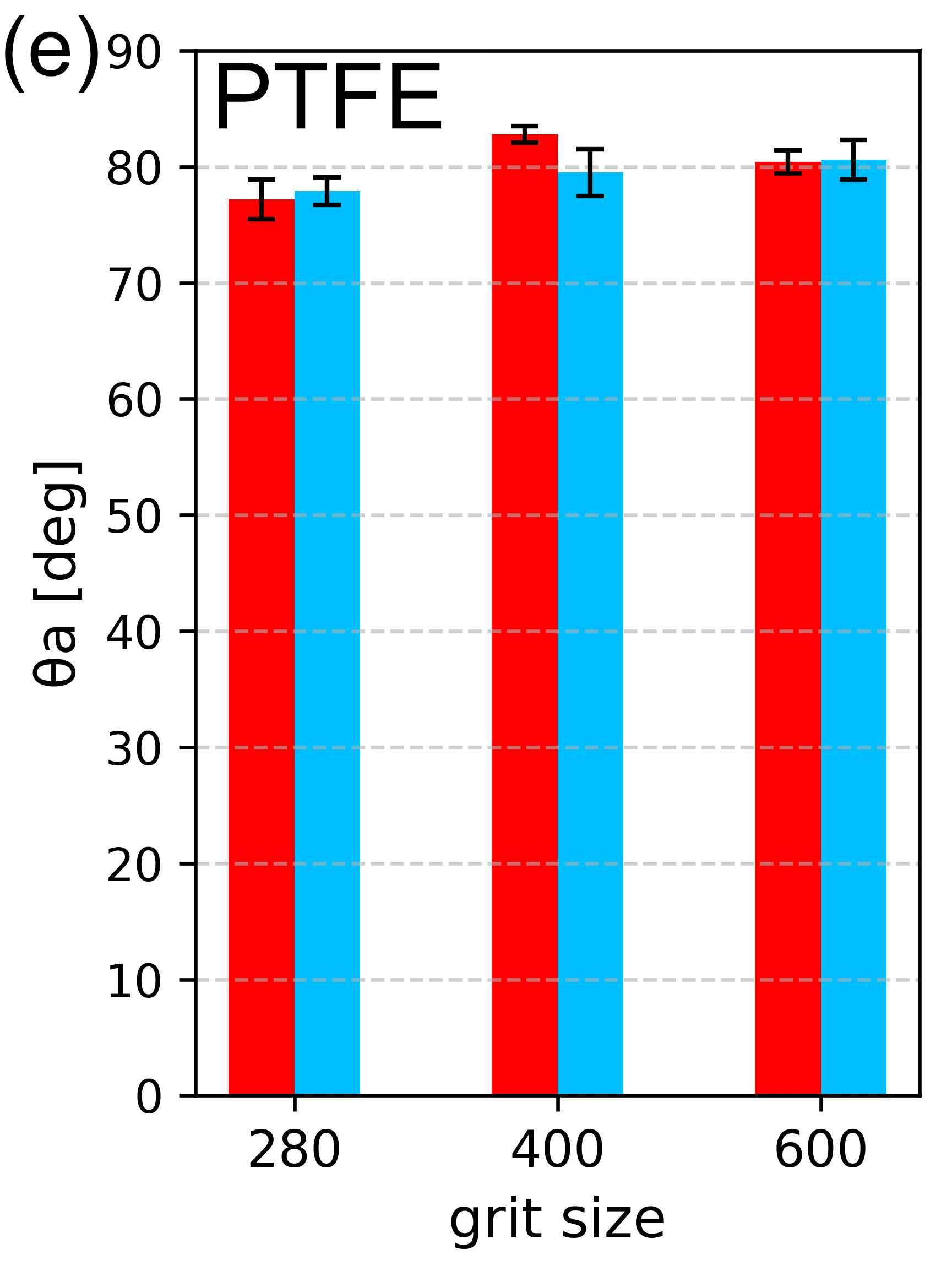}
\end{subfigure}
\hspace{1 mm}
\begin{subfigure}{0.3\linewidth}
\centering
\includegraphics[height=6 cm]{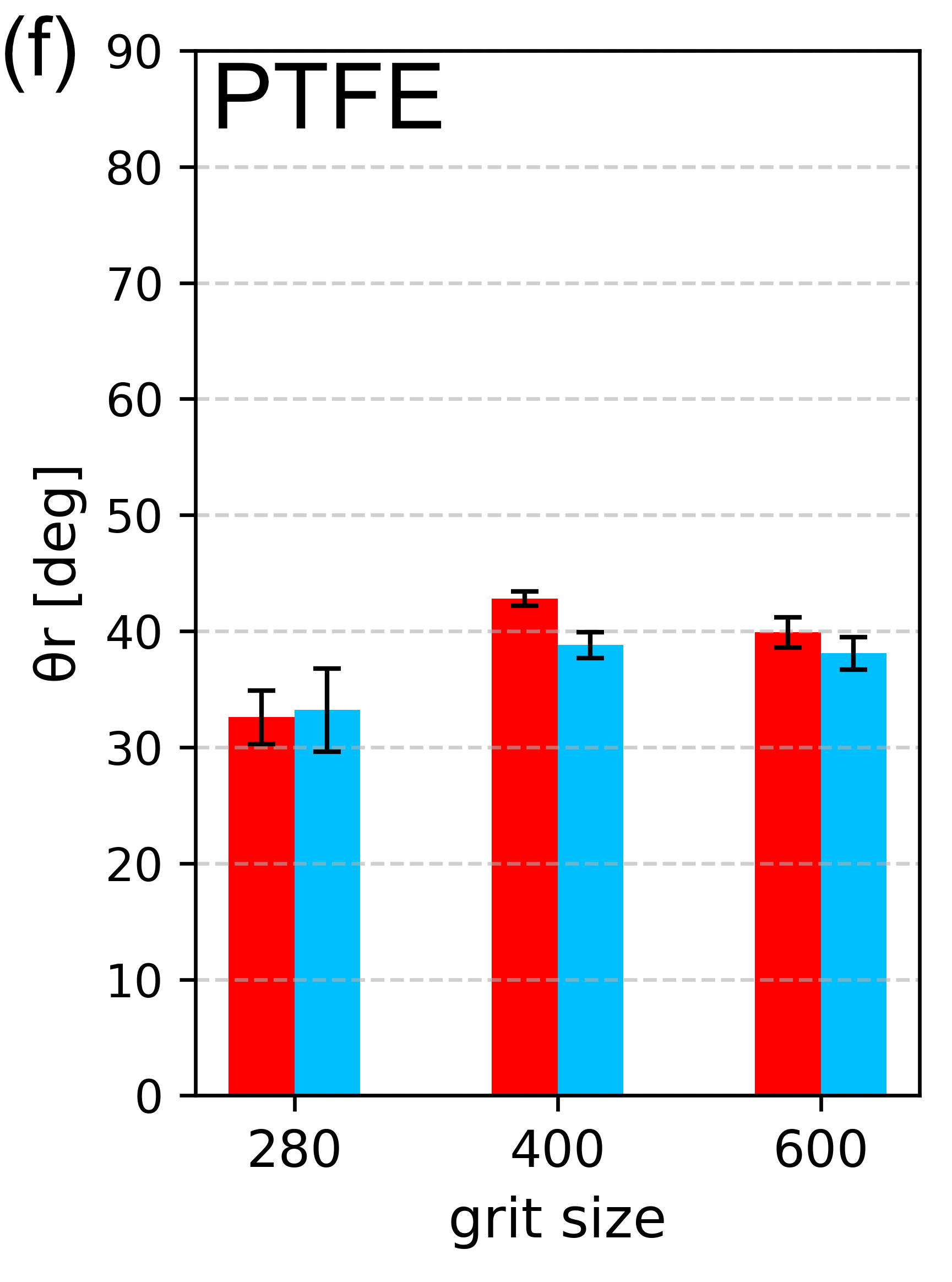}
\end{subfigure}

\caption{(a), (b) Dynamic contact angles measured on smooth polymer surfaces in air (droplets) and in water (bubbles). (a) Advancing contact angles. (b) Receding contact angles. (c - f) Dynamic contact angles measured on sandpaper-polished surfaces. (c) Advancing contact angle for PET. (d) Receding contact angle for PET. (e) Advancing contact angle for PMMA. (f) Receding contact angle for PMMA. Error bars represent standard deviations of three repeated runs.}
\label{results}

\end{figure}

\section{Results and Discussion}
We first present the results obtained using samples with smooth surfaces. As a representative example, Fig. \ref{PETresults} shows (a) a bubble during pushing, from which the receding contact angle is determined, (b) a bubble during pull causing the contact line advancing, from which the advancing contact angle is obtained, and (c) a droplet just before sliding, from which both the advancing and receding contact angles are determined on a PET plate. In addition, Fig. \ref{PETresults}(d) shows the time evolution of the bubble contact angle on a PET plate, where the horizontal axis represents time and the vertical axis represents the angle. The moment when the bubble first attached to the surface was defined as time zero. Under all experimental conditions, the time variation of the contact angle exhibited a step-like shape. The initial plateau corresponds to the stage in which the bubble was pushed against the surface. The upward-sloping region corresponds to the stage in which the bubble was being pulled away from the surface while the contact line remained pinned. The final plateau corresponds to the stage in which the contact line was advancing along the surface. The contact angle changed in a well-defined and stable step-like manner, allowing the advancing and receding contact angles to be distinguished unambiguously. The clear plateau regions and smooth transition between them indicate that the contact line behavior was stably captured with minimal fluctuation. The proposed method demonstrated good repeatability under the experimental conditions investigated in this study. The receding contact angle was defined as the average value during the final 20 s of the pushing process, whereas the advancing contact angle was defined as the average value during the first 20 s of the contact line advancing process. When the distance between the syringe and the surface becomes large, the bubble is stretched from the surface and its shape becomes significantly deformed, making accurate determination of the contact angle difficult in the image analysis.

Fig. \ref{results}(a) and (b) show the dynamic contact angles of bubbles and droplets measured on smooth polymer surfaces. The red bars represent the dynamic contact angles of bubbles measured in water using the modified captive bubble method, while the blue bars represent those of droplets measured in air. For all smooth surfaces, the advancing and receding contact angles obtained in water were in close agreement with those measured in air, with the error bars (standard deviations of three runs) overlapping in most cases. This result is consistent with previous studies reporting that dynamic contact angles measured in air and water are essentially identical on smooth surfaces\cite{Marmur1998}. More importantly, the scatter of the bubble measurements was consistently small, with the standard deviation ranged from \SI{0.7}{\degree} to \SI{2.2}{\degree}.

The edge-detection error, contact line position identification error, fitting error, and propagated error are summarized in Table \ref{error}. To evaluate optical distortion, an image of a grid with an approximately 1 mm spacing was captured, and the grid spacing was measured at the center and edge of the image. The measured spacings were 1.00 and 1.02 mm at the center and edge, respectively. Since the bubbles and droplets were imaged near the center of the lens, optical distortion was considered negligible and was therefore assumed not to affect the analysis. The stability of the bubbles and the influence of dissolved gas concentration were considered to have a negligible effect on the experimental results based on several observations. First, the bubble volume showed little change between before and after each experiment. In addition, because the bubbles used in this study were relatively large, minor changes in bubble volume were not expected to substantially affect their stability or the experimental results. Furthermore, a new bubble was generated for each trial, minimizing the potential influence of changes in bubble volume or gas exchange during repeated measurements. Finally, no fine bubbles were observed on the sample surface, as would be expected under conditions where the dissolved gas concentration is close to saturation\cite{Li2023}. Based on these observations, the dissolved gas concentration in the water was considered unlikely to have a significant influence on bubble stability or the experimental results.

\begin{table}[t]
\caption{The edge-detection error, corresponding to a threshold variation of 90±10; the contact line position identification error, corresponding to a variation of 20±10 data points in the distance from the slope sign-change position; the fitting error; and the propagated error.}
\centering
\begin{tabular}{l c c c c}
\hline
Sample & Edge detection & Contact line position identification & Fitting & Propagated \\
 & (deg) & (deg) & (deg) & (deg)\\
\hline
PET & 0.2 & 1.8 & 0.1 & 1.9\\
PC & 0.4 & 1.0 & 0.1 & 1.1\\
PMMA & 0.1 & 0.8 & 0.1 & 0.8\\
PTFE & 0.5 & 1.0 & 0.2 & 1.1\\
\hline
\end{tabular}
\label{error}
\end{table}

Next, Fig. \ref{results}(c) and (d) show the dynamic contact angles of bubbles and droplets for PET plates polished with three types of sandpaper, (e) and (f) show those for polished PMMA plates. When the dynamic contact angles in air and water were compared for surfaces with roughness, the results were also similar. For surfaces that exhibit the Wenzel state in air and the reversed gas–liquid Wenzel state in water, the dynamic contact angles are expected to be approximately identical in both environments because the liquid phase completely penetrates the surface roughness in each case. The present results are consistent with this expectation and demonstrate that the modified captive bubble method can reliably evaluate dynamic contact angles even for rough surfaces. The standard deviation ranged from \SI{0.4}{\degree} to \SI{2.3}{\degree} for bubbles.

As with the smooth surfaces, the scatter in the measurements on rough surfaces was also small. Previous studies have measured dynamic contact angles using conventional captive bubble methods on hydrophilic and hydrophobic smooth surfaces as well as on rough hydrophilic and hydrophobic surfaces, and reported the standard deviation ranging from \SI{2.4}{\degree} to \SI{5.2}{\degree}\cite{Xue2014}. In contrast, the standard deviations obtained in the present study were sufficiently smaller than these reported values. These results demonstrate that the modified captive bubble method provides a reliable approach for evaluating surface wettability in aqueous environments, with good repeatability under the experimental conditions investigated in this study.

Compared with the smooth PET and PMMA samples shown in Fig. \ref{results}(a) and(b), the dynamic contact angles of the polished samples were smaller. In particular, the receding contact angle decreased more significantly than the advancing contact angle. This behavior can be interpreted as a combination of two effects: (i) the enhancement of wetting when the roughness of a hydrophilic surface increases according to the Wenzel relation near the contact line\cite{Bico2001}, and (ii) the increase in local energy barriers caused by the pillars, which enhance the pinning effect acting on droplets and bubbles\cite{Gao2006}. As a result, both advancing and receding contact angles decrease due to enhanced wetting, while pinning effects increase the advancing contact angle and decrease the receding contact angle, leading to the observed behavior. Although the Wenzel state was not directly observed, the prepared rough surfaces were considered to be in the Wenzel state based on several observations. These included the surface roughness ($R_a$), the findings reported in the previous study\cite{Sarkar2021}, the lower contact angles compared with those of the corresponding smooth surfaces, and the random nature of the surface roughness. Furthermore, the fact that similar trends were obtained in water and in air supports the interpretation that the surfaces exhibited the inverted Wenzel state in water. Unlike the smooth surfaces, there were several cases in which the error bars did not overlap for the rough surfaces. However, the differences were not substantial. This discrepancy is likely attributed to the spatial variability of surface roughness introduced by manual sandpaper polishing, which results in random variations in surface morphology across the sample\cite{Montes2011}. In addition, the regions where the dynamic contact angles were measured were not identical for droplets and bubbles. These factors are considered to contribute to the observed deviations. On smooth surfaces, the effect of surface roughness on contact line pinning is expected to be negligible, and thus significant stick-slip motion is not observed. On rough surfaces, however, stick-slip motion may occur as a result of contact line pinning at surface asperities or microstructures. The present experimental setup, however, does not provide sufficient spatial and temporal resolution to directly resolve such microscopic contact line motion. Accordingly, we assume that any stick-slip motion does not significantly affect the macroscopic contact line behavior and use time-averaged contact angles in the present analysis. Further investigation of the contact line dynamics is an important subject for future work. In particular, magnifying the contact line region and using a high-speed camera would allow the microscopic motion of the contact line, including possible stick-slip behavior, to be directly observed and quantified.

\begin{figure}[t]
\centering

\begin{subfigure}{0.3\linewidth}
\centering
\includegraphics[width=4.5 cm]{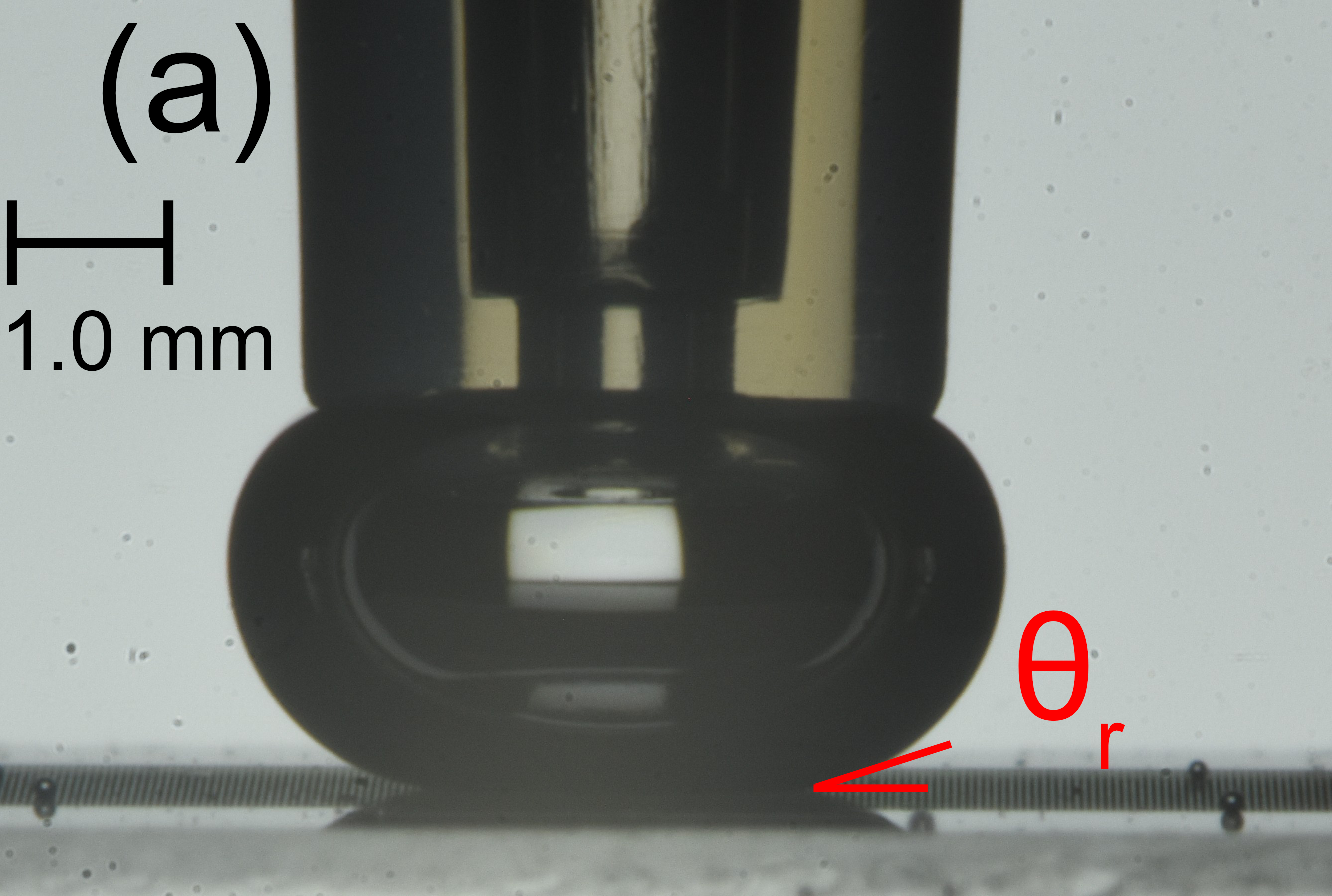}
\end{subfigure}
\hspace{1 mm}
\begin{subfigure}{0.3\linewidth}
\centering
\includegraphics[width=4.5 cm]{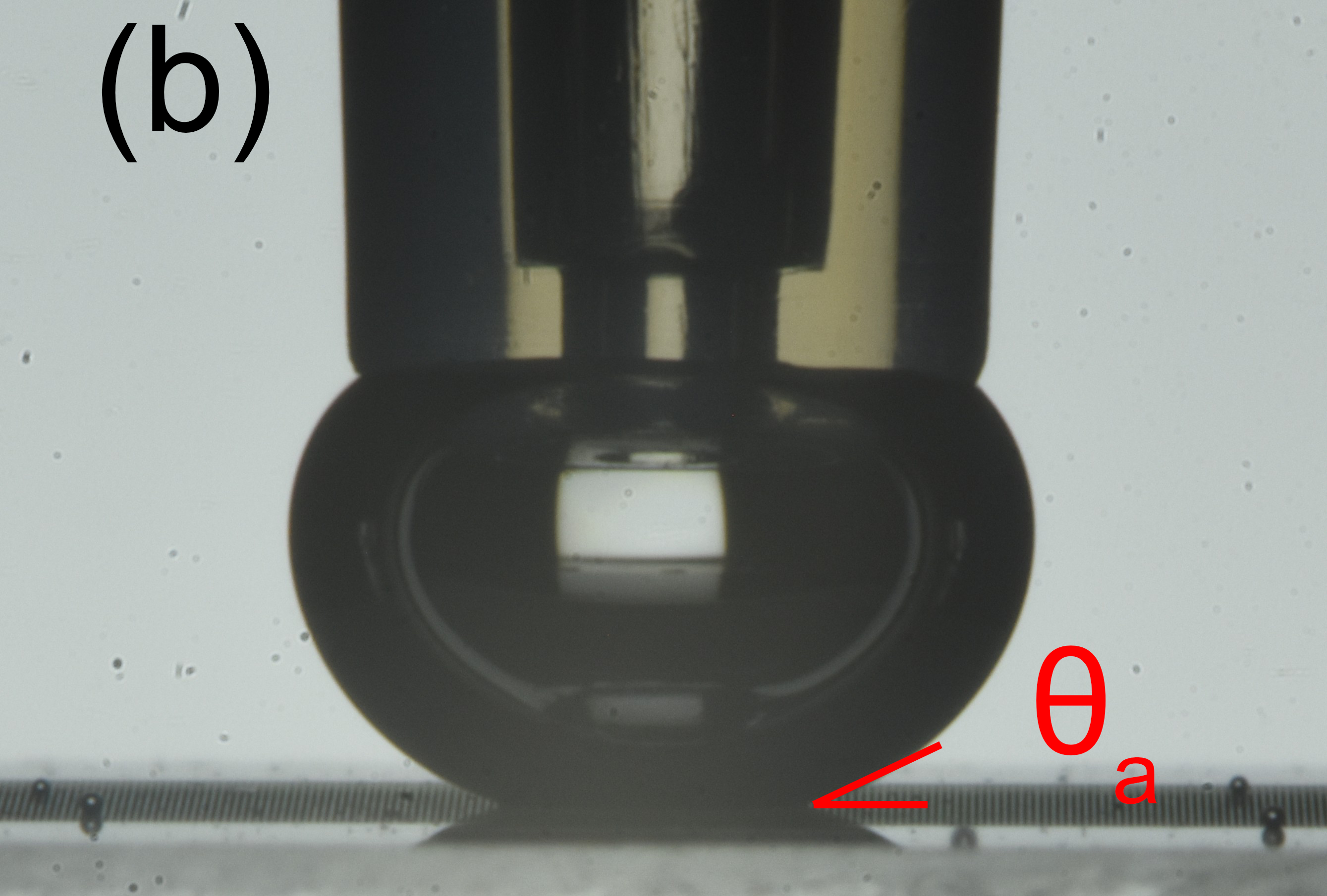}
\end{subfigure}
\hspace{1 mm}
\begin{subfigure}{0.3\linewidth}
\centering
\includegraphics[width=4.5 cm]{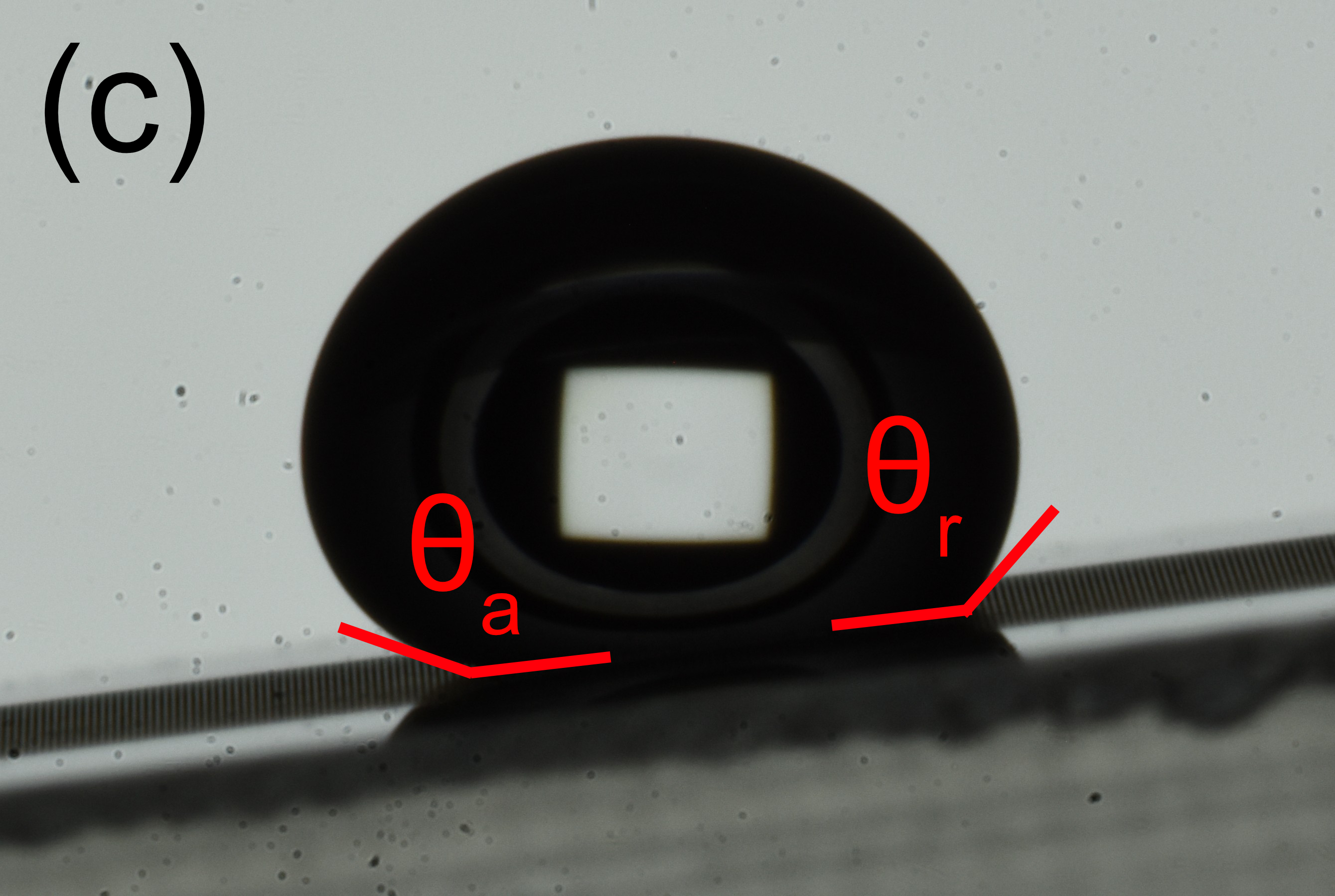}
\end{subfigure}

\vspace{1 mm}

\begin{subfigure}{1.0\linewidth}
\centering
\includegraphics[height=6 cm]{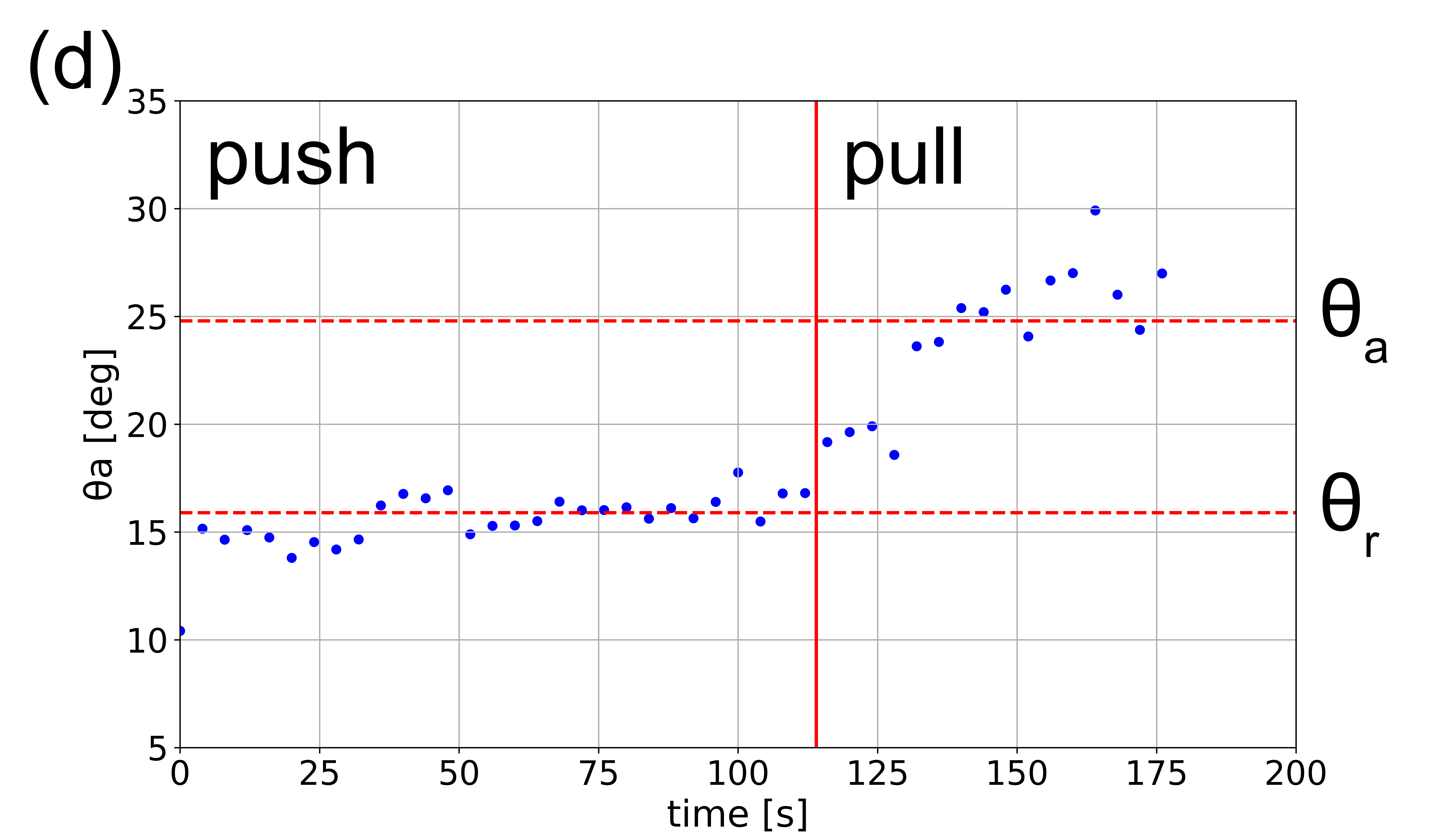}
\end{subfigure}
\caption{Measured data of dynamic contact angles on a microstructured PMMA surface. (a) Receding contact angle of a bubble during pushing against the surface. (b) Advancing contact angle of a bubble during the contact line advancing. (c) Advancing and receding contact angles of a water droplet measured in air just before sliding. (d) Time variation of dynamic the contact angle of bubble during the pushing and pulling processes of a bubble on a surface.}
\label{hydrophobic PMMA results}
\end{figure}

\begin{figure}[t]
\centering
\begin{subfigure}{0.3\linewidth}
\centering
\includegraphics[width=5 cm]{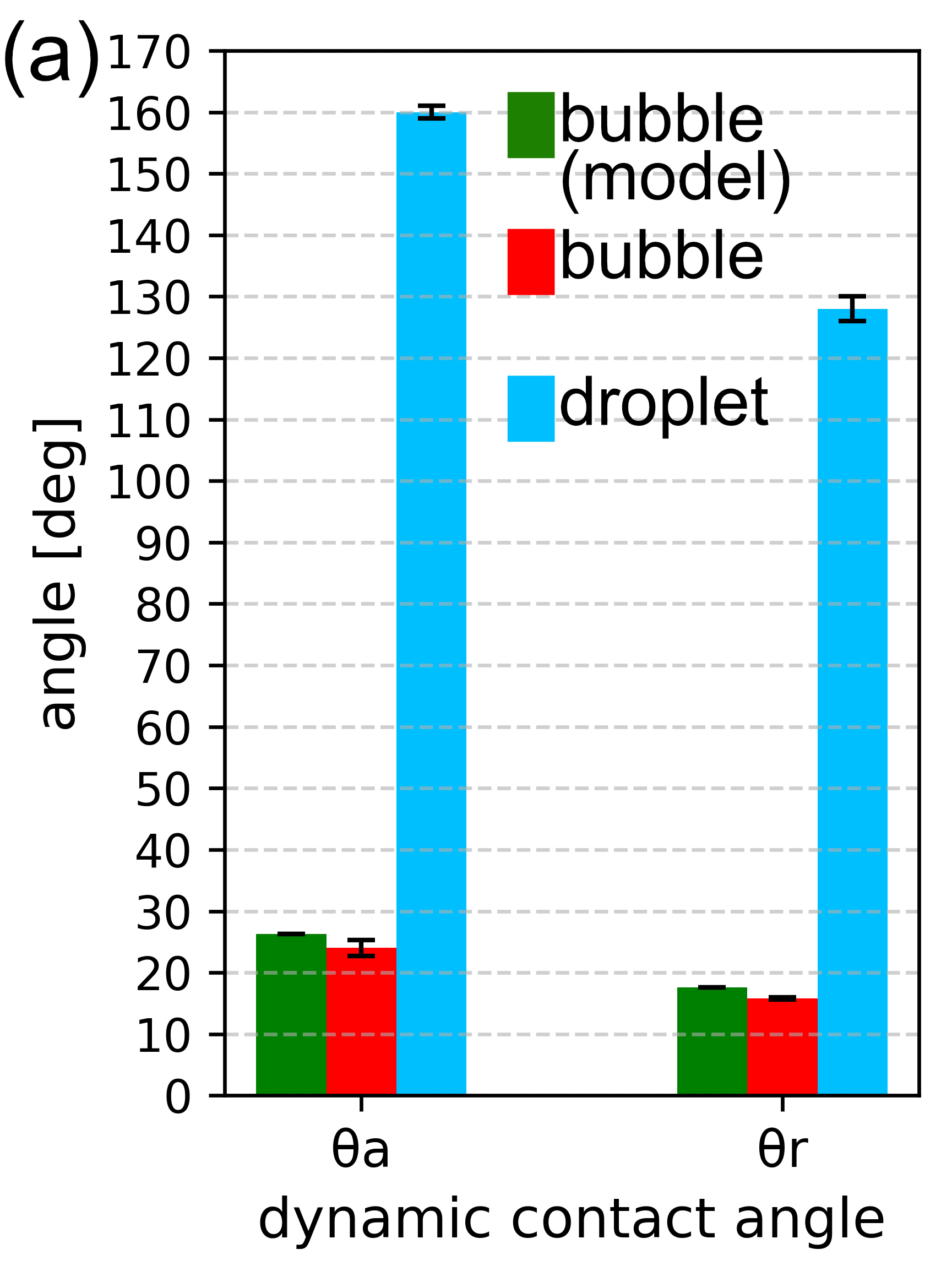}
\end{subfigure}
\hspace{1 mm}
\begin{subfigure}{0.3\linewidth}
\centering
\includegraphics[height=3.5 cm]{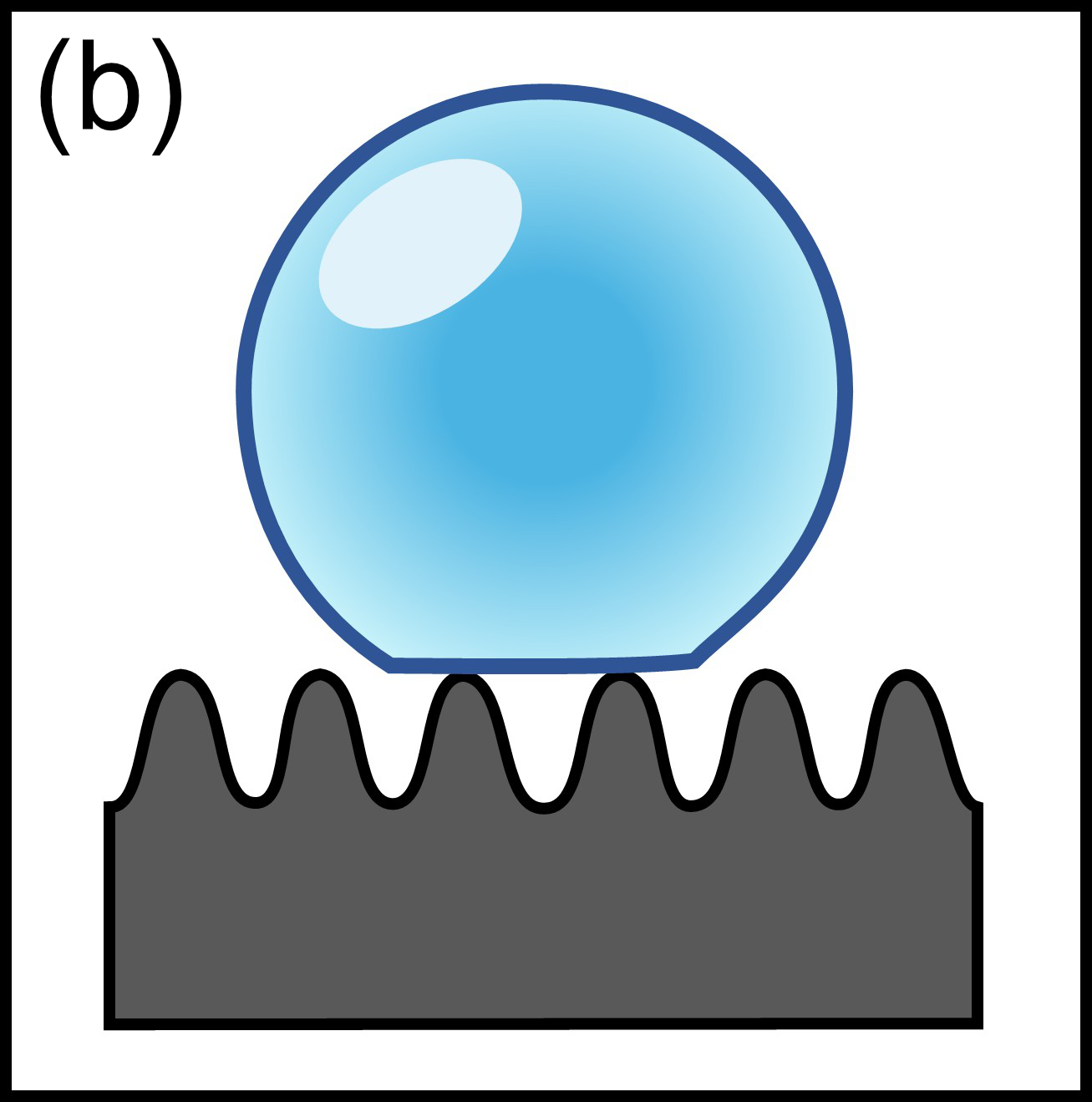}
\end{subfigure}
\hspace{1 mm}
\begin{subfigure}{0.3\linewidth}
\centering
\includegraphics[height=3.5 cm]{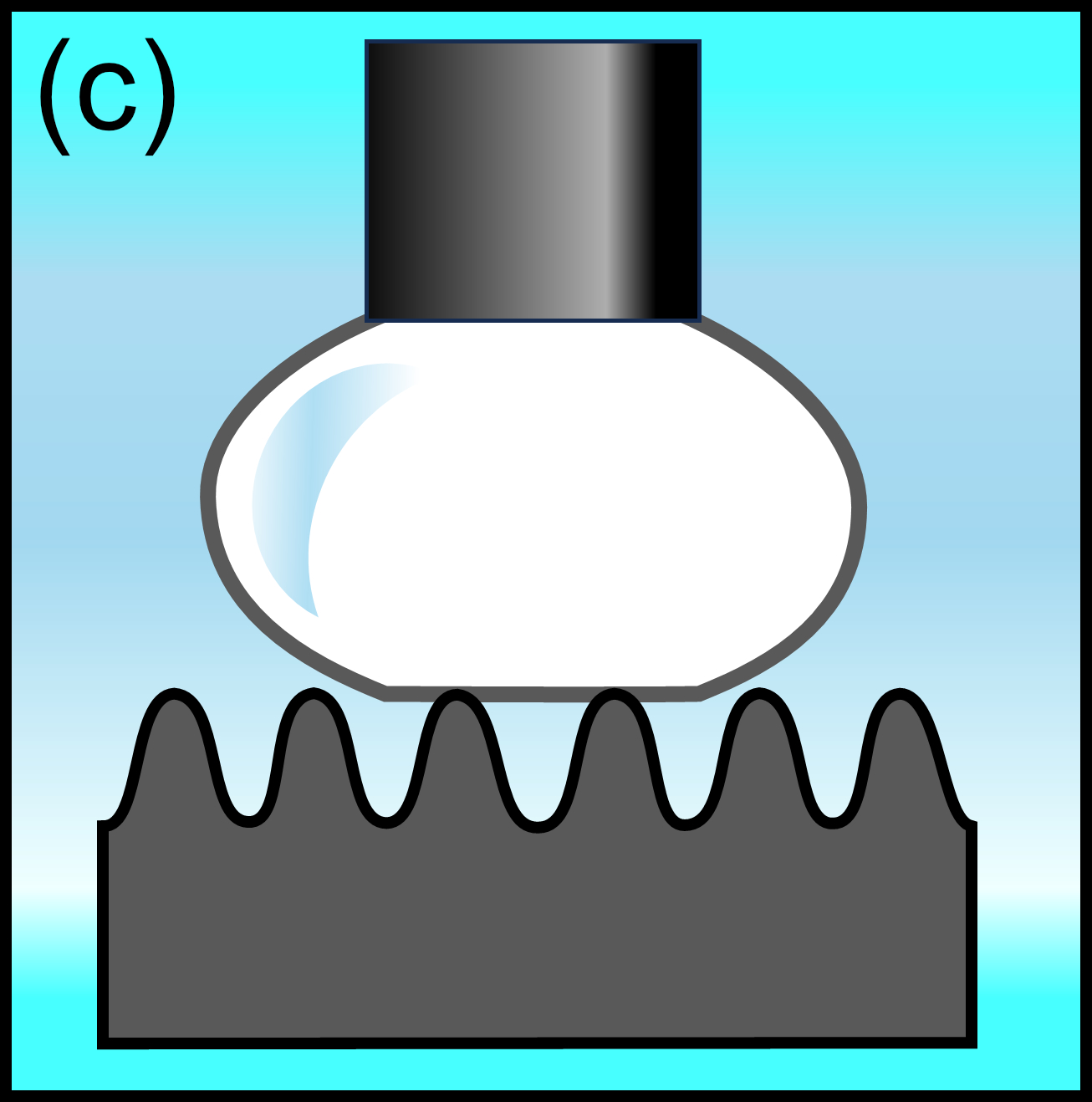}
\end{subfigure}
\caption{(a) Dynamic contact angles measured on hydrophobic PMMA surfaces in air (droplets), in water (bubbles), and that calculated using model in water. (b) Cassie-Baxter state in air. (c) Reversed gas–liquid Cassie-Baxter state in water.}
\label{Cassie}
\end{figure}

Finally, Fig. \ref{hydrophobic PMMA results} shows (a) a bubble during pushing, from which the receding contact angle is determined, (b) a bubble during the contact line advancing, from which the advancing contact angle is obtained, and (c) a droplet just before sliding, from which both the advancing and receding contact angles are determined on a microstructured PMMA plate as schematically shown in Fig. \ref{Wenzel}(c). As can be seen in comparison with Fig. \ref{PETresults}(a)–(c), the hydrophobic PMMA surface exhibits weak adhesion to both bubbles and droplets, indicating that they are easily repelled. In addition, Fig. \ref{hydrophobic PMMA results}(d) shows the time evolution of the bubble contact angle on a hydrophobic PMMA plate. Fig. \ref{Cassie}(a) shows the results of measured dynamic contact angles obtained using hydrophobic PMMA plates. When the dynamic contact angles in air and water were compared for this surface, the results differed significantly. The surface exhibited strong hydrophobicity in air but superhydrophilic behavior in water.

We discuss the reason for the significant difference in contact angles between droplets and bubbles. Assuming that the bubble did not penetrate into the grooves but instead contacted the protrusions of the PMMA structure and the surrounding water, corresponding to an inverted Cassie–Baxter state (Fig. \ref{Cassie}(b)(c)), we applied the Cassie–Baxter equation\cite{Cassie1944}:
 \[f_1 \mathrm{cos}\theta_1 + f_2 \mathrm{cos}\theta_2 = \mathrm{cos}\theta \]
In the Cassie–Baxter model, the apparent contact angle is determined by the area fractions of the different phases in contact with the liquid (or gas) at the interface. Here, \textit{$f_1$} and \textit{$f_2$} represent the area fractions of the solid surface and the secondary phase (e.g., air or liquid) in contact with the droplet or bubble, respectively, with \textit{$f_1$} + \textit{$f_2$} = 1. The angles \textit{$\theta_1$} and \textit{$\theta_2$} denote the intrinsic contact angles on each corresponding phase. This formulation is applicable not only to droplets in air but also to bubbles in liquid by appropriately defining the contacting phases. Although the contact angle is determined by the behavior of the contact line rather than the contact area\cite{McCarthy2007}, the length of the contact line in this experiment was much larger than the pillar pitch. The very small contact angle hysteresis observed for the bubbles suggests that the pinning effect caused by surface asperities is not significant. Therefore, the fraction of the contact line contacting the pillars was assumed to be approximately equal to the area fraction. Substituting \textit{$f_1$} = 0.09, which is estimated from the geometrical parameters of the surface (a square pillar top of \SI{15}{\micro\meter} and a pitch of \SI{50}{\micro\meter} (Fig. \ref{Wenzel}(c))), \textit{$f_2$} = 0.91 satisfying \textit{$f_1$} + \textit{$f_2$} = 1, and \textit{$\theta_2$} = \SI{0}{\degree}, based on the definition of the contact angle of a bubble on a liquid phase, the apparent contact angles were calculated. The use of \textit{$\theta_2$} = \SI{180}{\degree} in the Cassie–Baxter equation for a water droplet in air is based on the assumption that the liquid over air pockets is considered to be in a perfectly non-wetting condition with a \SI{180}{\degree} contact angle\cite{Kwok1999}. Following the same rationale, \textit{$\theta_2$} = \SI{0}{\degree} was used in the present inverted Cassie–Baxter model for a bubble immersed in water. During the laser fabrication process, the laser beam does not directly irradiate the top surfaces of the pillars. In addition, the temperature of the pillar surfaces is not expected to reach a level high enough to significantly alter their surface properties. Therefore, we consider that the surface properties of the pillar tops, which are in contact with the bubble in the proposed model, remain essentially unchanged. Consequently, the effect of any potential modification of the surface properties during laser processing on the present experimental results is considered to be negligible. By substituting \textit{$\theta_1$} = \SI{98.2}{\degree}, corresponding to the advancing contact angle of bubbles on smooth PMMA, \textit{$\theta_a$} was calculated to be \SI{26.3}{\degree}. Similarly, by substituting \textit{$\theta_1$} = \SI{62.2}{\degree}, corresponding to the receding contact angle, \textit{$\theta_r$} was calculated to be \SI{17.6}{\degree}. These calculated values are shown as green bars in Fig. \ref{Cassie}(a). The calculated values are close to the experimental results. It has been reported that when surfaces exhibiting the Cassie–Baxter state in air are immersed in water, a thin air layer often remains trapped within the microstructures, making accurate captive bubble measurements difficult or even impossible\cite{Sarkar2021}. Several attempts have been made to remove this trapped air layer; however, these approaches often cause damage to the surface structures\cite{Sarkar2021}. In this study, this problem was successfully resolved by ultrasonic degassing of the samples after immersion in water, which removed the trapped air without damaging the microstructured surface (Fig. \ref{sonic}). As a result, reliable dynamic contact angle measurements could be performed under fully wetted conditions. The results show that the dynamic contact angles in air differ significantly from those measured in water. We acknowledge that the present inverted Cassie–Baxter model is highly simplified. In addition, it was not possible to directly confirm whether ultrasonic degassing completely removed the air trapped within the microstructures. Therefore, the possibility that some residual air remained within the microstructures cannot be completely excluded. Nevertheless, the experimental results show very good agreement with the values predicted by the model, supporting the consistency of the model with the present experimental observations. At least, these results suggest that the amount of residual air was insufficient to significantly interfere with the bubbles brought into contact with the sample surface. Earlier studies have shown that rough randomly structured surfaces can exhibit different dynamic contact angles in air and water, and the present results are likely due to similar wetting behavior\cite{Sarkar2021,Xiao2022}. These findings indicate that surfaces exhibiting the Cassie–Baxter state in air show significantly different dynamic contact angles in air and water, and that the modified captive bubble method yields results consistent with measurements in air while showing good repeatability for smooth surfaces and for rough surfaces exhibiting the Wenzel state in air and the reversed gas–liquid Wenzel state in water.

\section{Conclusion}
In this study, we verified that the modified captive bubble method is an effective approach for measuring dynamic contact angles in aqueous environments. The main findings are summarized as follows. (i) The modified captive bubble method enabled stable observation of the contact line by suppressing bubble deformation and lateral displacement during the measurement process. (ii) For smooth surfaces and for rough surfaces exhibiting the Wenzel state in air and the reversed gas–liquid Wenzel state in water, the proposed method provides good repeatability for measuring underwater contact angles. A direct comparison with conventional captive bubble methods was not performed in this study, and therefore the relative accuracy and precision of the two methods remain to be investigated. (iii) For surfaces exhibiting the Cassie–Baxter state in air, ultrasonic degassing successfully removed the trapped air layer, enabling dynamic contact angle measurements using bubbles without damaging the surface structures. The measured values differed significantly from those obtained in air, demonstrating that a surface exhibiting the Cassie–Baxter state in air can exhibit the inverse Cassie–Baxter state in water.

This method expected to be useful for investigating adhesion phenomena between bubbles and solid surfaces and for the development of technologies that exploit such bubble adhesion. For example, the adhesion behavior of bubbles to submerged particles can only be evaluated by observing whether particles remain attached to a bubble when the bubble is pressed against and subsequently detached from the particle surface\cite{Albijanic2010}. In addition, this work may have potential applications in fields such as marine science, where surface wettability influences biofouling and organism attachment\cite{Epstein2010}, and medicine, where interfacial properties affect biological adhesion and the performance of biomaterials\cite{Sun2021}.

%
%

\ack{The authors gratefully acknowledge Hikari Kikai Seisakusho Co., Ltd. for preparing the microstructured PMMA plate used in this study.}

\funding{This work was supported by JSPS KAKENHI Grant number JP25K22010 and JST ERATO Number JPMJER2401.}

\roles{Koki Iwasaki: Conceptualization, Methodology, Investigation, Formal analysis, Visualization, Writing – original draft. Hiroyuki Ebata: Formal analysis, Writing – review and editing, Supervision. Hiroaki Katsuragi: Conceptualization, Methodology, Writing – review and editing, Supervision, Funding acquisition.}

\data{The data supporting the results of this study are available within the article. Supporting datasets have been deposited in Zenodo, including raw data underlying the results, the corresponding graphs for each sample, and numerical data used to generate the graphs. Each experiment was repeated three times for each sample; representative data from a single measurement are provided in the repository. The data are available at Zenodo: https://doi.org/10.5281/zenodo.19059902}

\conflicts{There are no conflicts to declare.}

\Ethics{This study did not involve human participants or animals.}


\bibliographystyle{unsrt}
\bibliography{References}

\end{document}